\documentclass[floatfix,aps,prb,twocolumn,groupedaddress]{revtex4}
\usepackage{graphicx} 
\usepackage[]{epsfig}
\usepackage[]{psfig}
\usepackage{times}
\usepackage[abs]{overpic}
\newcommand{\beq}{\begin{equation}}
\newcommand{\eeq}{\end{equation}}
\newcommand{\bea}{\begin{eqnarray}}
\newcommand{\eea}{\end{eqnarray}}
\newcommand{\nn}{\nonumber}
\newcommand{\no}{\noindent}

\newcommand{\eps}{\epsilon}
\newcommand{\veps}{\varepsilon}
\newcommand{\al}{\alpha}
\newcommand{\s}{\sigma}
\newcommand{\lam}{\lambda}
\newcommand{\de}{\delta}
\newcommand{\D}{\Delta}

\newcommand{\be}{\beta}

\newcommand{\vp}{\varphi}
\newcommand{\ra}{\rangle}
\newcommand{\la}{\langle}

\newcommand{\ga}{\gamma}
\newcommand{\om}{\omega}
\newcommand{\ka}{\kappa}

\newcommand{\tht}{\theta}
\newcommand{\app}{\approx}

\newcommand{\imag}{{\rm Im}}
\begin{document}
\bibliographystyle{apsrev}
\title{Kondo screening cloud effects in mesoscopic devices}
\author{Pascal Simon}
\email{pascal.simon@grenoble.cnrs.fr}
\affiliation{Department of Physics and Astronomy, University of Basel, CH-4056 Basel, Switzerland \\
 LEPES \& LPM2C, CNRS, 25 av. des martyrs, 38042 Grenoble, France}

\author{Ian Affleck}
\email{affleck@physics.bu.edu}
\affiliation{Physics Department, Boston University, 590 Commonwealth Ave., 
Boston, MA02215}
\date{\today}
\begin{abstract}
We study how finite size effects may appear when a quantum dot in the Kondo Coulomb blockade regime is embedded into a mesoscopic device with finite wires. These finite size effects appear when
the size of the mesoscopic device containing the quantum dot is of the order of the
size of Kondo cloud and affect all thermodynamic and transport properties of the Kondo quantum dot. We also generalize our results to the experimentally relevant case where the wires
contain several transverse modes/channels. Our results are based on perturbation theory, Fermi liquid theory and slave boson mean field theory.
\end{abstract}
\maketitle

\section{Introduction}
In the last years, a large number of experiments have been realized in order to probe 
some theoretical predictions related to the Kondo effect. This has been made possible
using small semiconductor quantum dots \cite{dot,Cronenwett,Wiel} and also thin single wall carbon nanotubes.\cite{Cobden}
In  these devices, it is expected that when the ground state of the quantum
dot or of the thin carbon nanotubes has a spin
$S=1/2$
or equivalently when
its last full level contains on average  one 
electron and becomes degenerate, the dot
plays the role of the Kondo impurity when embedded into conducting leads.

One of the major and still controversial issue in the Kondo physics concerns
the Kondo screening cloud. The heuristic 
picture is that an electron  in an extended wave function with a size of order $\xi_K=\hbar v_F/T_K$ (the Kondo length scale) surrounds the impurity, forming a singlet with it. The remaining electrons outside the cloud do not feel  the impurity spin at low energies but rather a  scattering potential, 
caused by the complex formed between 
the cloud and the impurity, resulting in a $\pi/2$ phase shift
at the Fermi energy (some recent experiments by Heiblum {\it et al} \cite{Heiblum} have tried to measure this
$\pi/2$ phase shift but obtain results in disagreement with theoretical predictions.\cite{Costi} Nevertheless,
the issue of  the eventual relation between the phase experimentally measured and the 
$\pi/2$ phase shift is still controversial \cite{unitarity}).
The Kondo temperatures in the experiments on quantum dots are generally 
  smaller than $1^0K$ and  most importantly can
be tuned via the gate 
  voltage $V_g$. Therefore,  the Kondo length scale is expected to be 
 very large. Typically with a Fermi velocity $v_F=5.10^5 m/s$, $\xi_K$ is  of order 3 microns.
Such low tunable Kondo temperature may  therefore
offer new opportunities to find evidence of  this screening cloud. 
	
Recently, we have proposed that persistent currents in a ring containing a quantum dot
may offer a way to probe directly this large Kondo length
scale. \cite{Affleck}( See also Refs 
[\onlinecite{Simon,Hu,Johan,Kang1,Ferrari,Aligia,Comment,Sorensen}]).
When the Kondo screening cloud becomes of order of the ring size or larger, we expect
a crossover from a (large amplitude) saw tooth shape  to a (small amplitude) sinusoidal shape.
Unfortunately, persistent currents experiments are generally delicate and
very sensitive to disorder. In order to probe this fundamental length scale,
it has been highly 
desirable to have some theoretical predictions based on conductance measurements which are more commonly used and better controlled.

Possible devices to probe the Kondo length scale   might be a quantum dot embedded 
into or connected to a finite size geometry.
Several geometries where some finite size effects due to the large Kondo length scale may occur have been proposed recently: possible geometries include a quantum dot
connected to
a box,\cite{Thimm,Balseiro} a mesoscopic ring \cite{Kang} or a quantum dot embedded into
short wires. \cite{Letter,Cornaglia}
The device we want to study in detail in this paper consists of a quantum dot embedded
in two short wiress (of length comparable to the Kondo length scale) which are then weakly coupled to some reservoirs. Such device is schematically depicted in Figure 1.
One may wonder whether the conductance is sensitive to whether the Kondo cloud
leaks into the reservoirs or is trapped inside the short leads.
In the latter case, we may expect the conductance to be mainly dominated by the contacts between the reservoirs and the short leads. The answer is not so simple in the former case. 
Some partial results about this geometry have already been published elsewhere. \cite{Letter}
We have indeed studied a symmetric geometry, assuming the wires contain only one transverse channel, 
by means of perturbative calculations combined with a Fermi liquid approach when available.  
We have shown that when the bare Kondo temperature $T_K^0$ 
(defined by the Kondo temperature for infinite wires) is of the order or smaller than the energy level spacing
in the wires, then the genuine Kondo temperature associated with  the impurity can be very different from $T_K^0$
and strongly depends on the local density of states seen by the dot being on or off resonance.
In this paper, we would like to provide a more detailed analysis of this geometry and discuss the generality
of the predicted finite size effects. 
The plan of the paper is the following: in section 2, we present the tight binding model
we want to study and give detailed derivations
of these finite size effects using perturbation theory in the Kondo couplings.
In section 3, we extend these results to wires which contain several transverse modes
or channels. We note that the case of infinite length multi-channel wires has never been 
fully discussed to the best of our knowledge.
In section 4, we show how these finite size effects affect quantitatively the transport properties through the quantum dot.
In section 5, we present complementary results valid at low temperature
obtained with the Slave Boson Mean
Field Theory (SBMFT). Finally in section 6, we discuss the main approximations
made and the generality of our results. Three appendices with technical details
are at the end of the paper.

\section{Finite size effects in a box geometry} 
\subsection{Model Hamiltonian}

In this section, we want to study the following system:
 a quantum dot embedded in a quantum wire which 
is in turn 
connected to external leads by weak tunnel junctions.    We assume 
that a gate voltage can be applied to the dot and also to the quantum 
wires.
A related device has been proposed recently by Thimm
{\it et al.} \cite{Thimm} where the quantum dot modeled by a standard Anderson Hamiltonian
 was  coupled to all energy levels
of a finite size  box with the same (energy independent) tunneling amplitude. 
The energy level spacing was assumed constant and 
of $O(1/V)$ where $V$ is the volume of the (3 dimensional) box. 
These two aspects differ considerably from the geometry studied here,
since the electrons need to pass through the
dot to contribute to the conductance. In fact, in the geometry studied in 
[\onlinecite{Thimm}], the box is ``side-coupled''
to the dot whereas, here, the box embeds the dot. 
Moreover, the Non Crossing
Approximation used in [\onlinecite{Thimm}] might be questionable in such
geometry where several new energy scales emerge compared to the usual Kondo model as was shown in [\onlinecite{Letter}].
\begin{figure}
\epsfig{figure=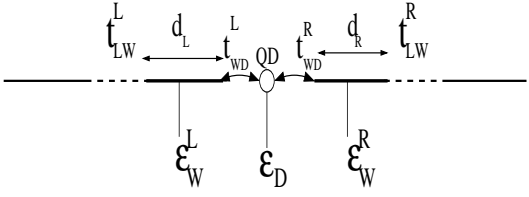,height=2.5cm,width=7cm}
\caption{Schematic representation of the device under consideration. $\eps_D$
  and $\eps_W^{L/R}$ control respectively the dot and wires gate voltage.}
\label{device}
\end{figure}
 We consider a  one-dimensional tight-binding model which describes 
our device and has the Hamiltonian:
\begin{equation}
\label{ham}
H=H_L+H_W+H_D+H_{LW}+H_{WD},
\end{equation} 
where the subscripts $L$, $W$ and $D$ stand for leads, wires and dot respectively.  Here:
\begin{eqnarray}
H_L&=&-t\left[ \sum_{j=-\infty}^{-d_L-2}+\sum_{j=d_R+1}^{\infty}\right]
(c^\dagger_jc_{j+1}+H.c.)\nonumber \\
H_W&=&-t\left[ \sum_{j=-d_L}^{-2}+\sum_{j=1}^{d_R-1}\right]
(c^\dagger_jc_{j+1}+H.c.)\nn   \\
&&+\epsilon_W^L \sum_{-d_L}^{-1}n_j+
\epsilon_W^R\sum_{1}^{d_R}n_j  \\
H_D&=&\epsilon_Dn_0+Un_{0\uparrow}n_{0\downarrow}\nonumber \\
H_{LW}&=&-(t_{LW}^Lc^\dagger_{-d_L-1}c_{-d_L}
-t_{LW}^R c^\dagger_{d_R}c_{d_R+1}+H.c.)\nonumber \\
H_{WD}&=&-(t_{WD}^Lc^\dagger_{-1}c_0+t_{WD}^R c^\dagger_0c_1+H.c.)\nn
\label{ham1}
\end{eqnarray}
Here $n_{j\sigma}\equiv c^\dagger_{j\sigma}c_{j\sigma}$ and 
$n_j\equiv n_{j\uparrow}+n_{j\downarrow}$. Note that $t_{WD}^i$ and $t_{LW}^i$, with $i=L,R$ (this notation will be used throughout the paper), 
are the  hopping amplitudes
between the different sections of the geometry under consideration (see Figure \ref{device}).
The spin indices have been omitted in order to lighten notations.

This Hamiltonian is a very simplified version of the geometry of interest in this paper.
We assume that the quantum wires are in the ballistic regime and do not contain 
 any other impurities (magnetic or not).
We  first consider the case of $1$ channel where partial results have been reported elsewhere. \cite{Letter}
We will also discuss further a more realistic modeling of the quantum wires which typically contain $5$ to $10$ 
transverse channels \cite{Wiel} at the end of this section. 
We  ignore electron-electron interactions 
in the wires and leads, only keeping them in the dot. 
This point will be discussed in  section \ref{discuss}.

We  assume that the system is in the strong Coulomb blockade regime, 
so that $t_{WD}<<-\epsilon_D$, $U+\epsilon_D$, where $\epsilon_D<0$.  
Then we may eliminate the empty and doubly occupied states of the 
dot, so that $H_D+H_{WD}=H_K$ gets replaced by a Kondo interaction
plus a potential scattering term:
\begin{eqnarray}\label{defhk}
H_K&=&2J(\ka_L c^\dagger_{-1}+\ka_R c^\dagger_1)
{\vec \sigma \over 2}(\ka_Lc_{-1}+\ka_Rc_1)\cdot \vec S \nn\\
&+&2V(\alpha_Lc^\dagger_{-1}+\alpha_Rc^\dagger_1)(\alpha_L c_{-1}+\alpha_R c_1)
\end{eqnarray}
with 
\bea
J&=&(~(t_{WD}^L)^2+(t_{WD}^R)^2)\left[ {1\over -\epsilon_D}+{1\over U+\epsilon_D}\right]\\
V&=&{(t_{WD}^L)^2+(t_{WD}^R)^2\over 4}\left[ {1\over -\epsilon_D}-{1\over U+\epsilon_D}\right]
\eea
and
\beq
\ka_R={t^R_{WD}\over \sqrt{(t_{WD}^R)^2+(t_{WD}^L)^2}} \ \;\ \ \ka_L={t^L_{WD}\over \sqrt{(t_{WD}^R)^2+(t_{WD}^L)^2}}
\eeq

Here $\vec S$ is the spin operator for the quantum dot.

The Kondo effect can be understood as resulting from a renormalization 
of the Kondo coupling constant, $J$, to large values at low temperatures.  
Perturbation theory is infrared divergent but the temperature acts 
as an infrared cut-off yielding a finite result which is accurate 
if the temperature is sufficiently high ($T>>T_K$).  At low temperatures, 
a non-perturbative description is needed.  This is provided by the 
local Fermi liquid description.\cite{Nozieres}  If we imagine that 
$J>>t$, then 
a spin singlet forms in the groundstate from the impurity and an electron 
in a symmetric orbital on sites $\pm 1$.  The anti-symmetric orbital 
still remains available to conduct current so the system is roughly 
equivalent to the $U=0$ model and exhibits resonant conductance with 
the resonance tied to the Fermi surface. \cite{Simon}

In the case 
of a closed ring, we showed earlier that the renormalization of the Kondo
coupling 
is cut off, even at low temperatures, by the ring circumference.\cite{Affleck}  In 
the present situation, this renormalization would be cut off by the finite 
length, $L=d_L+d_R$, of the quantum wires, if $t_{LW}=0$.  Essentially, if the 
Kondo cloud doesn't have enough space to form, then the growth 
of the Kondo coupling constant is cut off. The effective 
Kondo coupling at the length scale $L$ becomes of O(1) when 
$L\approx \xi_K$. Nevertheless, the situation is not so simple for   
 small but finite $t_{LW}$.  
The growth 
of the Kondo coupling is no longer cut off by the finite size of the wire.
Nonetheless, some noticeable finite size effects are expected to occur when 
$L$ is  reduced to a value of $O(\xi_K)$, which corresponds to the screening 
cloud beginning to leak into the leads.

\subsection{Kondo temperature of the system}\label{Kondotemp}

Before computing any thermodynamic or transport properties of the system, it is crucial to
know the genuine Kondo temperature of our system, $T_K$, the main energy scale of the problem.
In order to calculate $T_K$, we first diagonalize the Hamiltonian 
at $J=0$, i.e. we diagonalize 
$H_0\equiv H_L+H_W+H_{LW}$.
Since the sites 
at $j<0$ and $j> 0$ are completely decoupled in this limit, 
we can diagonalize the parts corresponding to the right and left leads separately.
Let us first focus on the left part only. 
If $t_{LW}^L=0$, then the wave-functions 
and eigenvalues of the left part (corresponding to $-d_L\leq j\leq -1$) are:
\begin{eqnarray} 
\psi_L(j)&=&(1/\sqrt{d_L})\sin k_{L,n}j\nonumber \\
k_{L,n}&=&\pi n/(d_L+1);~~1\leq n\leq d_L \\ 
\veps(k_{L,n})&=& -2t\cos k_{L,n}+\epsilon_W^L-\mu\nn.\end{eqnarray}  
For non-zero $t_{LW}^L$, the spectrum of $H_0^L$  (the left part of $H_0$ is continuous).
In order to study how the Kondo interaction renormalizes, 
we express $c_{-1}$ in terms of the  eigenstates, 
$c_\epsilon$ of $H_0^L-\mu N_L$:
\beq
c_{-1} = \int_{-2t-\mu}^{2t-\mu} d\epsilon f_L(\epsilon)c_{L,\epsilon}.
\eeq
We have normalized the operators $c_{L\eps}$ so that $\{c_{L,\eps}^\dagger,c_{L,\eps'}\}=\delta(\eps-\eps')$.
The left local density of states is defined by $\rho_L(\eps)=|f_L(\eps)|^2$
and is normalized according to $\int_{-2t-\mu}^{2t-\mu} d\epsilon \rho_L(\eps)=1.$
This local density of states can be easily computed exactly for our tight binding model.
The calculation  of $\rho_L(\eps)$ can be found in the appendix \ref{density}.
The final result, assuming $\ga_L=t_{LW}^L/t\ne 0$ is
\begin{widetext}
\beq
\rho_L(k)={1\over \pi t} {\ga_L^2\sin^2 k_L\sin k\over \sin^2 k_L(d_L+1)-2\ga_L^2\cos k\sin (k_Ld_L)\sin(k_L(d_L+1))+\ga_L^4\sin^2 (k_Ld_L)}
\label{rhol}
\eeq
where $k_L$ is related to $k$ by 
\beq \label{defkn}
-2t\cos k-\mu=-2t\cos k_L +\eps_W^L-\mu.
\eeq
The fine structure and the properties of this local density of states has been studied in the appendix \ref{density}.
For small $t_{LW}^L$, this local density of states, 
$\rho_L (\epsilon )\equiv |f_L(\epsilon )|^2,$
has sharp peaks at the energies
$
\epsilon_n=\epsilon_W-\mu -2t\cos [k_{L,n}],
$
where the momenta $k_{L,n}$ are solutions of the equation:
\beq
2k_L(d_L+1)+\arctan{{\ga_L\sin(k+k_L)\over 1-\ga_L^2\cos(k+k_L)}}-\arctan{{\ga_L\sin(k-k_L)\over 1-\ga_L^2\cos(k-k_L)}}\label{kln}
=2\pi n,
\eeq
with the constraint between $k$ and $k_L$ given by Eq. (\ref{defkn}).
For small $\ga_L$, the solution of this equation is approximately
\beq
k_{L,n}\app \pi n/(d_L+1)+O(\ga_L^2/d_L).
\eeq
Plugging this value of $k_{L,n}$ in (\ref{defkn}) provides the value of the gate
voltage $\eps_W^L$ necessary to reach such a resonance.
The separation between two consecutive peaks  is approximately 
$\Delta_{L} \approx 2\pi \sin(k_{L,n})/d_L$.
The 
width of these peaks can be also evaluated as 
\beq
\delta_{L,n} ={2(t_{LW}^L)^2\sin^2(k_{L,n})\sin(k)\over t(d_L+1)}.\label{width}
\eeq
We can deduce that the ratio of width to separation  for a peak defined by $k_{L,n}$ is of order:
\begin{equation}
\delta_{L,n} /\Delta_{L,n} \app {t^2_{LW}\over t^2}{\sin(k_{L,n})\sin(k)\over \pi}.
\end{equation}
We will assume that this quantity is $<<1$, a condition already encountered for $\ga_L\approx 0.5$.    
Obviously, similar results are obtained for the right part of $H_0$ by changing $L\to R$.

The full Hamiltonian
may be written in this basis as:
\bea
&&H-\mu N_L-\mu N_R=\int_{-2t-\mu}^{2t-\mu} d\epsilon \epsilon
 (c_{L,\epsilon}^\dagger c_{L,\epsilon}+c^\dagger_{R,\epsilon} c_{R,\epsilon})\nn\\
&& + \int d\epsilon d\epsilon{'}
\left\{  f_L^*(\epsilon )f_L(\epsilon{'})
\left(J_{LL}c^\dagger_{L,\epsilon}{\vec \sigma \over 2}\cdot \vec Sc_{L,\epsilon'} +V_{LL} c^\dagger_{L,\epsilon}c_{L,\epsilon}\right)\right.
+  f_R^*(\epsilon )f_R(\epsilon{'})
\left(J_{RR}c^\dagger_{R,\epsilon}{\vec \sigma \over 2}\cdot \vec S c^\dagger_{R,\epsilon'}+V_{RR}c^\dagger_{R,\epsilon}c_{R,\epsilon'}\right)\nn\\
&&+\left.\left( f_L^*(\epsilon )f_R(\epsilon{'})(J_{LR}c^\dagger_{L,\epsilon}{\vec \sigma \over 2}\cdot \vec Sc_{R,\epsilon'}+ V_{LR})c^\dagger_{L,\epsilon}c_{R,\epsilon'}+H.c.\right)\right\}
\label{HE}
\eea
where we have defined  $J_{LL}=2\kappa_L^2 J,J_{RR}=2\kappa_R^2 J,J_{LR}=2\kappa_L\kappa_R J$. Similar definitions hold for the couplings $V$. It is worth noticing that these notations should not hide the fact that there is only one genuine Kondo coupling $J$ and therefore only 
one Kondo temperature. Nevertheless, these notations are useful for transport properties
as we be seen below.

Using the Hamiltonian in the form of 
Eq. (\ref{HE}), one can study how the Kondo couplings renormalize when the band-width is lowered from $\pm D_0$ 
(where $D_0$ is O($t$)] to $\pm D$. At second order in the Kondo couplings, we obtain:
\beq \label{RG}
J_{ij}\to J_{ij}+{1\over 2}\sum_k J_{ik}J_{jk} 
\left[ \int_{-D_0}^{-D}+\int_{D}^{D_0}\right] {d\epsilon \rho_k (\epsilon )
\over |\epsilon |},
\eeq
\end{widetext}
where $i,j,k=L$ or $R$.
The renormalization of the Kondo couplings  is quite different depending on how far we 
lower the cut off, $D$.  If $D>>\Delta_{i,n}$ ($i=L,R$),  the integrals in Eq. (\ref{RG}) 
average over many peaks in the densities of states so their detailed 
structures become unimportant and we obtain the result for the usual Kondo
model (see for example [\onlinecite{Glazman}]).
In particular, for a completely symmetric geometry, we recover the standard result
$J\to J[1+2J\rho_0\ln (D_0/D)]$ ($J=J_{LL}=J_{RR}=J_{LR}$).

On the other hand, for smaller $D$, 
$D<<\Delta_{i,n}$, the renormalization of $J$ in Eq. (\ref{RG}) becomes strongly dependent 
on the fine structure of the local densities of states and therefore of $\epsilon_W^L$ and $\epsilon_W^R$. Let us analyze this in details in the next subsections. 

\subsubsection{Fully symmetric geometry}
\begin{figure}
\epsfig{figure=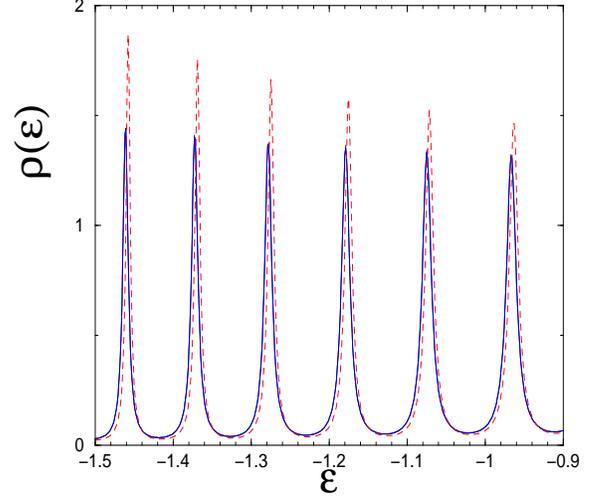,height=7cm,width=7.5cm}
\caption{The exact (plain blue curve) and the approximate (dashed red curve) local densities of states as a function of $\eps$. We have taken $d=d_L=d_R=49$ and $\ga=0.5$.}
\label{densfig}
\end{figure}

In order to simplify the picture, we first consider the completely symmetric case.
In this subsection, most of the subscripts $L$ and $R$ have been suppressed.
Let us first assume that $\epsilon_W$ is tuned to 
a resonance of the density of states of width $\delta_n$.  
Then the integral in Eq. (\ref{RG}) 
gives a very small contribution as $D$ is lowered from $\Delta_n$ down 
to $\delta_n$ so the renormalized Kondo coupling at the scale $D$, $J_{eff}(D)$, practically stops renormalizing over 
this energy range.  Finally, when $D<\delta_n$, the density of states 
grows rapidly.
To go further, we can approximate
 the  local density of states (at the left or right part of the dot) by a sum of several Lorentzians of 
width $\de_{n}$:
\beq
\label{ldosapprox}
\pi\rho_i(\eps)\approx{2\over d_i}\sum\limits_{n=1}^d \sin^2k_{i,n} {\de_{n}\over (\eps-\eps_n)^2+\de_{n}^2},
\eeq
where $k_{i,n}=\pi n/(d_i+1)+O(\ga^2/d_i)$ ($i=L=R$). This constitutes a quite good approximation
when $\ga^2\ll 1$. For example we have plotted in figure \ref{densfig} the exact result and the approximate result for $\ga=0.5$. We see that both curve agree remarkably well even for $\ga=0.5$. They differ slightly in the height of the peaks and in the peak positions especially at the edges of the band.

We can  express the result in terms 
of the change in $J_{eff}(D)$ as $D$ is lowered from $\Delta_n$ around the resonance peak at $\eps=\eps_n$:
\begin{widetext}
\begin{equation}
J_{eff}(D)\approx J_{eff}(\Delta_n )[1+J_{eff}(\Delta_n ){4\sin^2(k_{i,n})\over \pi d \delta_n}
\ln ({\delta_n\over D})],\label{jeff}
\end{equation}
\end{widetext}
where $d=d_L=d_R$.
When $\eps_W$ is tuned exactly on resonance (i.e. $\eps_W=\eps_n$), 
the density of states appearing in this renormalization is enhanced by 
a factor of 
$(2t\sin^2 k_{i,n}/[d\sin k_n\delta_n])\approx t^2/[t_{LW}^2\sin^2 k_n]$,  $k_n$  being related to $k_{i,n}$ by (\ref{defkn}).
This leads to 
a rapid growth of $J_{eff}(T)$. 

On the other hand, if $\epsilon_W$ is 
off-resonance then the density of states is small, of order 
$(\delta_n /\Delta_n^2~d)$ so the growth of the Kondo coupling is 
very slow at all energies $D<\Delta_n$. 

It is clear that this difference in the renormalization process between the two
 cases will strongly affect
the Kondo temperature of the system. We
define the Kondo temperature as the temperature where $J_{eff}(T)$ becomes of $O(1)$.  
When
$J_{eff}(T)$ becomes large at $T>>\Delta_n$ then $T_K$ is related to the 
bare Kondo coupling and bandwidth as in the usual case (with no weak links):
$T_K\approx T_K^0\equiv D_0e^{-1/2J\rho_0}$. Furthermore, in this case, 
$T_K$ does not depend strongly on $\epsilon_W$.  We may characterize 
this case by $T_K^0>>\Delta_n$ or equivalently $\xi_K<<L$.  The screening 
cloud fits inside the quantum wires and the weak links do not modify 
the Kondo effect significantly.  

\begin{figure}
\vskip 0.5cm
\includegraphics[height=7cm,width=8cm]{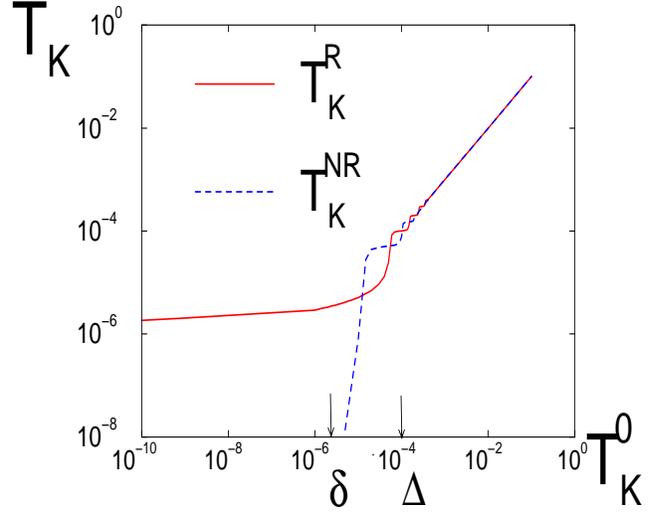}
\caption{
The curves represent $T_K=f(T_K^0)$ 
in a log-log scale keeping the same values for $\D_n, \de_n$ for $\eps_W$ on
resonance (plain curve which becomes almost flat at low $T_K^0\ll \de_n$),
and $\eps_W$ off resonance (dashed curve which drops sharply at low $T_K^0$).
Both curves coincide at $T_K^0>\D_n$. 
}\label{tkeff}
\end{figure}

On the other hand, suppose that $T_K^0<<\Delta_n$ implying that
 $J_{eff}(\Delta_n )<<1$.  In this case $T_K$ depends strongly on
 $\epsilon_W$.  If the system is tuned to a resonance then $T_K$ will 
be slightly less than $\delta_n$:
\beq
T_K^{R}\approx \de_n\left( {T^0_K\over
D_0}\right)^{t_{LW}^2\sin^2k_n/t^2}=O(\de_n),
\eeq
  On the other hand, if the system 
is off-resonance then $T_K<<\delta_n$, the Kondo temperature reads
\beq
T_K^{NR}\app \Delta_n\left({T_K^0\over D_0}\right)^{t^2/(t_{LW}^2\sin^2k_{d,n})}.
\label{tkor}
\eeq  
In this equation the subscript $NR$ stands for non resonant.
In general, for small values of $t_{LW}/t$, we expect $T_K^{NR}\ll T_K^{R}$ (for experimental purposes, 
$T_K^{NR}$ is almost $0$).
We have plotted the behavior of $T_K$ vs. $T_K^0$   in  
Figure \ref{tkeff} for both $\eps_W$ on resonance and off
resonance. 
This figure clearly illustrates the change of behavior when $T_K^0\gg \Delta_n$ or 
$T_K^0\ll \Delta_n$. 
The curves
coincides for $T_K^0\gg \D_n$ as expected. When $T_K^0\sim \Delta_n$, the Kondo temperature begins to be sensitive to the
fine structure in the local density of states which are signaled by the small oscillations of both $T_K^{R}$ and 
$T_K^{NR}$ (for the non resonant case, we have chosen the Fermi energy 
symmetrically between two resonant peaks) . Note that there is a regime ($\D/2<T<\D$) where $T_K^{OR}>T_K^{R}$. It corresponds to the situation where the integrals
in (\ref{RG}) are dominated by two peaks for the off-resonance case  and by one peak only for the on-resonance case.  
When
 $T_K^0\ll \D_n$
the off resonance Kondo temperature $T_K^{OR}$ drops sharply at $T_K^0<\D_n$
to very small values ($\ll \de_n$). On the other hand $T_K^R$  also has
a sharp drop at $T_K^0<\D_n$ but then becomes almost flat and of order $\de_n$.
Since the Kondo temperature, the main energy scale, enters in the calculation of 
all most all the  properties of the system, this strong difference of behavior 
between the on or off 
resonance case
will affect dramatically all transport or 
thermodynamic quantities.

\subsubsection{Non-symmetric geometry}
If we consider now the general case of a non symmetric geometry, $3$ different situations may occur
by tuning the two gate voltages $\eps_W^L$ and $\eps_W^R$: (i) they are both on resonances, (ii) both  off resonances, or (iii)
just one is on resonance. 
In case (i) or (ii), the physics will be very similar to what we discuss in the fully symmetric case.
A natural way to define the Kondo temperature is to evaluate $D$ such that  the ratios between 
the second order order corrections to the Kondo couplings and their bare values
$J_{LL}^{(2)}/J_{LL}^{(0)},J_{RR}^{(2)}/J_{RR}^{(0)},J_{LR}^{(2)}/J_{LR}^{(0)}$
equals one. Since all the Kondo couplings are coupled together, we expect this to occur
at the same value of $D=T_K$, the energy scale of the problem. Provided   the local 
densities of states are of the same order of magnitude, the discussion in the previous section can be repeated. 
New features emerge for case (iii). Let us suppose that $\eps_{W}^L$ is tuned on resonance and
$\eps_{W}^R$ off resonance. If $T_K^0$ is much larger than $max(\Delta^L,\Delta^R)$, we obviously 
recover the bulk situation with $T_K=T_K^0$. When $T_K^0$ is smaller than the level spacing, then 
$T_K$ will depend on the fine structure of the densities of states.
The RG equations (\ref{RG}) are then dominated by the resonance peak in the left lead.
It means that the impurity is essentially screened in the left part of the device.
Neglecting terms involving $\rho_R$ in the flow (\ref{RG}), the Kondo temperature associated with this hybrid situation (we use the notation $T_K^{H}$ for this hybrid case)
can be evaluated
as before as
\beq
T_K^{H}\approx \delta_n^L \left( {T_K^0\over D_0}\right)^{2(t_{LW}^L)^2\sin^2(k_n)/t^2}\approx (T_K^R)^2/\delta_n.
\eeq
We therefore also expect the Kondo temperature to be of order $\de_n^L$. Nevertheless, the strong asymmetry between the left and right part will affect transport properties.

\section{Extension to the multi-channel case}

So far, we have modeled the short wires by a $1$ dimensional tight binding model 
containing only 
$1$ transverse channel.
Nevertheless, the semiconducting wires contain in general $5$ to $10$ 
transverse modes.
In this section, we discuss extensions of the previous results to take into account 
this multi-channel situation. Before considering the general case of finite length
wires, we report for completeness on the more standard case
of infinite wires connected to a quantum dot.

\subsection{Case of infinite wires} \label{infinitewire}
\subsubsection{Tight binding formulation}
This situation can be easily taken into account by introducing in  our tight 
binding Hamiltonian formulation a channel index (or band index) $\al=1,\dots,N$ to the electron operators
which labels the electronic eigenmodes in the transverse direction.

The tight binding Hamiltonian for infinite wires  reads 
\begin{eqnarray}\label{ham2}
H &=&-t\left[ \sum_{j=-\infty,\al}^{-2}+ \sum_{j=1,\al}^{\infty}\right]
(c^\dagger_{j,\al}c_{j+1,\al}+H.c.) \nn\\
&-&\sum_\al (t^L_{WD,\al} c^\dag_{-1,\al}c_0+t^R_{WD,\al} c^\dag_{1,\al}c_0+H.c.)\nn\\
&-& \sum\limits_{\al,i} \mu_\al n_{i,\al}+\eps_D n_0+Un_{0,\uparrow}n_{0,\downarrow}
\end{eqnarray}
Note that the spin indices have been omitted.
It is straight forward to extend the analysis developed in the previous section.
After a Schrieffer-Wolff transformation, the Kondo interaction reads:
\bea \label{hkondo1}
H_K&=&\sum_{\al,\be=1}^N \left[J^{LL}_{\al,\be} c^\dag_{-1,\al} {\vec \sigma \over 2}
\cdot \vec S c_{-1,\be}+
J^{RR}_{\al,\be} c^\dag_{1,\al} {\vec \sigma \over 2}
\cdot \vec S c_{-1,\be}\right]\nn\\
&+&\sum_{\al,\be=1}^N  \left[J^{LR}_{\al,\be} c^\dag_{-1,\al} {\vec \sigma \over 2}
\cdot \vec S c_{1,\be}+H.c.\right].
\eea
with 
\beq \label{defJs}
J^{ij}_{\al,\be}=2t_{WD,\al}^it_{WD,\be}^j/\tilde \eps_D\eeq
with 
$\tilde \eps_D=\left(\frac{1}{-\eps_D}+\frac{1}{-\eps_D+U}\right)$.
In this section, we neglect the direct scattering terms since they turn out to be
unimportant for our discussion. 
In order to study the renormalization of this set of Kondo couplings,
it may be more convenient to momentarily suppress the distinction between left
and right lead and to directly work with $2N$ channels. This can be done
by $c_{-j,\al}\to c_{j,\al+N}$, $J^{LL}_{\al,\be}\to J_{\al+N,\be+N}$ and
$J^{LR}_{\al,\be}\to J_{\al+N,\be}$. We will switch between these
two notations throughout this section.
With these notations, the RG equations take a simple form
\beq \label{RGJ}
J_{\alpha \gamma}\to J_{\alpha \gamma}+
J_{\alpha \gamma}
\sum_{\beta}J_{\beta \beta}
\int d\epsilon {\rho_{\beta}(\epsilon )\over |\epsilon |},\eeq
where we have introduce $\rho_\beta(\eps)$ the local density of states seen by the dot
in the channel $1\leq \beta\leq 2N$. Note that the equation (\ref{RGJ}) follows
from the fact the Kondo coupling $J_{\al\be}$ can be written as a product
of a term related to channel $\al$ with a term related to channel $\be$. 
This product form is preserved under RG at least to lowest order.
One may also introduce the dimensionless Kondo couplings constants
\beq \label{deflambda}
\lambda_{\alpha,\beta}=\sqrt{\rho_\al\rho_\beta} J_{\al,\be}.
\eeq
Since we consider infinite wires, it seems reasonable to assume that the densities of states are constant ($\rho_\al=\sin(k_{F,\al})/\pi t$ in our tight binding model) in each channels such that the RG equations can be further simplified to
\beq \label{RGlambda}
{d\Lambda\over d\log l}=\Lambda^2+O(\Lambda^3),
\eeq
where $\Lambda=(\lambda_{\al,\be})$ is a matrix of Kondo couplings. As usual
 the RG equations make sense provided $\lambda_{\al,\be}^{(0)}\ll 1$.
All the couplings constant are driven to the strong coupling regime. Nevertheless,
there is only one scale in the problem. This can be seen by noticing that the
matrix $\Lambda^0$ (which defines the bare values of the Kondo couplings)
 has only one non zero eigenvalue ($=Tr \Lambda^0$). This is directly
related to the fact that only one ``effective'' channel couples at the boundary
to the impurity.
 Therefore, the Kondo temperature can be defined by:
\beq
T_K^0=D\exp\left[-1/Tr(\Lambda^{(0)}\right].
\eeq 
At $T\ll T_K^0$, the impurity is screened by this ``effective'' channel.
It is straightforward to extend the analysis of Ng and Lee \cite{Ng} (which relies
on the Langreth description of the low temperature Anderson model \cite{Langreth}) in order
to prove that the conductance through a quantum dot {\it symmetrically} 
connected to two identical leads (containing several channels) reaches the unitary limit $2e^2/h$.
In the more general case corresponding to   two leads non identically connected to the dot,
the $T=0$ conductance in the Kondo regime reads
\beq
G={2e^2\over h} {\sum\limits_{\al,\be=1}^N 4\Gamma_\al^L(\eps_F) \Gamma_\be^R(\eps_F) 
\over \left[\sum\limits_{\al=1}^N (\Gamma_\al^L(\eps_F)+\Gamma_\al^R(\eps_F) )\right]^2}
,\eeq
with $\Gamma^L_\al(\eps)=(t_{WD,\al}^L)^2\sin(k_{F,\al})/t  $.

\subsubsection{Continuum limit formulation}
It seems also interesting to look at this multichannel case 
in the continuum limit.
We have seen and concluded in the previous subsection 
that one ``effective'' channel
couples at the boundary to the impurity. One may therefore wonder 
whether we can reduce
the multi-channel problem to a single channel one.
In order to formulate this model in the continuum limit, we may 
 linearize the spectrum of the left/right leads 
around the $2N$ different Fermi points as:
\beq
c^{L/R}_\al(x)= c_{\al}(\pm x)=e^{-i k_{F,\al}} \psi_{R,\al}^{L/R}(x)+  e^{i k_{F,\al}}
\psi^{L/R}_{L,\al}(x)
\eeq
$~\forall ~x>0$ and 
where $\psi^{L/R}_{L,\al} $ represents a left moving electron in the left or right lead.
Instead of working with left and right movers living on a semi infinite line
we can work with left  movers only on the infinite line using the fact that we have perfectlyreflecting boundary conditions.
The linearized Hamiltonian reads:
\bea
H_{lin}&=&\int\limits_{-\infty}^{+\infty} dx \left(\sum\limits_{\al=1}^N 
\sum_{i=L,R}{v_\al^F\over 2\pi}
(\psi^i_{L,\al})^\dag(x) i\partial_x \psi^i_{L,\al}\right.\nn\\
&+&\left.{\delta(x)\over 2\pi}\sum\limits_{\al,i} [t^i_{\al} c_{d}^\dag \psi^i_{L,\al}(x)+H.c.]
\right)+H_{D},\label{hlin}
\eea
with $t^i_\al=2i\sin k_{F,\al} t^i_{WD,\al}$, $v_\al^F=2t\sin(k_{F,\al})$ and $i$ labels the left and right side of the dot.
Since the impurity couples only to one linear combination of operators $ \psi^i_{L,\al}(0)$, it 
may be tempting     to
introduce a new field basis 
defined by a unitary transformation $U^i$ to the field $\psi^i_{\al}$ defined by
\beq
\eta^i_{1}=\sum\limits_\al u_{1,\al} \psi^i_{\al}~~~{\rm with}~~~
u_{1,\ga}={t^i_\ga\over \sqrt{\sum\limits_\al |t^i_\al|^2}}.
\label{unitary}
\eeq
The other set of operators $\eta^i_{\al}$ with $\al>1$ can be built in such a
way for the matrix $U=(u)_{\al,\be}$ to be unitary by the Schmidt orthonormalization process.
Nevertheless, this unitary transformation leaves the kinetic part diagonal provided $v_\al^F=v^F~~\forall \al$.
If this condition is met,
the Anderson impurity model reduces in this new basis only to one effective channel. 
A similar conclusion was recently reached independently by Cho et al. \cite{Cho} 
considering several independent leads (containing one transverse channel).
Since at low temperature, transport  depends only on the Fermi level properties, 
the conductance through the dot will be dominated by
one
effective channel. In particular, for a symmetric geometry, we recover that
the conductance may reach at most the value
$2e^2/h$
corresponding to the unitary limit. 
Nevertheless, for several transverse modes in a wire the condition $v_\al^F= v_F$ is most likely not satisfied  and the introduction of the unitary transformation $U$ may be not very helpful since it does not leave the kinetic energy 
diagonal.
(One can still argue that the difference between Fermi velocities
is usually irrelevant in $1D$ systems and the conclusions obtained above should hold.)
Therefore, when the Fermi velocities are different, we need to perform
some RG analysis analogous to what was developed in the previous section.
In this formulation, the Kondo Hamiltonian reads:
\beq \label{hkcont}
H_{K} =\sum\limits_{\al,\be=1}^{2N} \sqrt{v^F_\al v^F_\be}\lambda_{\al,\be}
\psi^\dagger_{\al}
{\vec \sigma\over 2}\cdot S \psi_{\beta}
\eeq
where $\lambda_{\al,\be}$ has been defined in (\ref{deflambda}).
Note that we have used the compact notation with $2N$ channels in order 
to avoid the distinction between left and right wires. Starting from
Eq. (\ref{hkcont}), one can obtain in a similar manner 
 the RG equations given in Eq. (\ref{RGlambda})

\subsection{Case of finite size wires}

In this section, we  extend  our previous results to take into account 
the fact that the wires are now finite and contain several channels.
A possible Hamiltonian which extends Eq. (\ref{ham}) and describes this situation
 reads 
\beq  H=H_L+H_W+H_D+H_{LW}+H_{WD}+H_G\eeq
 with
\begin{eqnarray}\label{hamgen}
H_L&=&-t\left[ \sum_{j=-\infty,\al}^{-d_L-2}+\sum_{j=d_R+1,\al}^{\infty}\right]
(c^\dagger_{j,\al}c_{j+1,\al}+h.c.) \\
H_W&=&-t\left[ \sum_{j=-d_L,\al}^{-2}+\sum_{j=1,\al}^{d_R-1}\right]
(c^\dagger_{j,\al}c_{j+1,\al}+h.c.)\nn   \\
&&+\epsilon_W^L \sum_{j=-d_L,\al}^{-1}n_{j,\al}+
\epsilon_W^R\sum_{j=1,\al}^{d_R}n_{j,\al}  \\
H_{LW}&=&-\sum\limits_{\al,\be}(t_{LW,\al\be}^Lc^\dagger_{-d_L-1,\al}
c_{-d_L,\be}\nn \\&&
-t_{LW,\al\be}^R c^\dagger_{d_R,\al}c_{d_R+1,\be}+H.c.) \\
H_{WD}&=&-\sum_\al(t_{WD,\al}^Lc^\dagger_{-1,\al}c_0+t_{WD}^R c^\dagger_0c_{1,\al}+H.c.)
\\
H_G&=& -\sum\limits_{\al,i} \mu_\al n_{i,\al}.
\end{eqnarray}
$H_D$ is the dot Hamiltonian and remains unchanged compared to \ref{ham}.
Note that the tunneling amplitudes between the wires and the leads 
mix together the channels. Allowing channel mixing will not change much our discussion as we will see further.
The different bands have  dispersion relations
\bea \label{defknal}
\veps_\al(k)&=&-2t\cos k-\mu_\al\nn\\
&=& -2t\cos k_L +\eps_W^L-\mu_\al\\&=&-2t\cos k_R +\eps_W^R-\mu_\al\nn.
\eea

\subsubsection{Without channel mixing}
Before treating the general case of a matrix of tunneling amplitudes
between the wires and the leads, we first consider in this subsection the 
simpler case where the matrices $t_{WL}^{L/R}$ are diagonal in the channel 
index.
Such an  approximation is valid  when the wires and the 
leads are geometrically 
similar (i.e. with the same number of tranverse modes) and the tunneling barrier separating them 
has no dependence on the transverse direction. 

The fact that $t_{LW,\al}\ne t$ does not modify much the discussion made in section
 \ref{infinitewire}. A Schrieffer-Wolff transformation leads to the same Kondo
Hamiltonian given by Eq. (\ref{hkondo1}). The main difference is that the local density of states seen by the quantum dot acquires
some non trivial energy dependence.
Therefore, the RG equations corresponding to Eq. (\ref{RGJ}) are still valid.
The density $\rho_\al(\eps)$ generalizes Eq. (\ref{rhol}) with $\ga_L\to \ga_{\al}=
t_{WL,\al}/t$, $k_L\to k_{\al}$, $\mu\to \mu_\al$ and corresponds to the LDOS in channel $1\leq \al\leq 2N$ (within this compact notation $t_{WL,\ga}^R=t_{WL,\ga},~ t_{WL,\ga}^L=t_{WL,\ga+N}$ $\forall \ga\in[1,N]$). 
Note that there is still only one scale in the problem, the Kondo temperature 
$T_K=D\exp[-/Tr(\Lambda{(0)}]$, which depends on the local density of states
seen by the impurity, more exactly on the total tunneling amplitude
$\Gamma(\eps)=\pi\sum_\al t_{WD,\al}^2 \rho_\al(\eps)$ (see Eq. (\ref{deflambda})).

The properties of $\rho_{\al}$ are therefore similar
to those of $\rho_L(\eps)$ studied in section \ref{Kondotemp}.
Let us for example  analyze the total left tunneling amplitude seen by 
the quantum dot
defined by $\Gamma_{L}(\eps)=\pi\sum\limits_{\al=N+1}^N t_{WD,\al}^2 \rho_{\al}(\eps)$.
Since the $\mu_\al$ are different, the position of the peaks in the
LDOS $\rho_{L,\al}$ are in general different.
It implies that the {\it average} level spacing in the quantity $\Gamma_{L}(\eps)$ is now
of order $\D^L_N=\pi v_F/Nd_L\app \D_L/N$. Moreover the peaks have now different widths
depending on $t_{WL,\al}^L$, $\mu_\al$ and $\eps_W^L$. The analysis developed 
in  section \ref{Kondotemp} extends straightforwardly: when $T_K^0$ becomes of order
of or smaller than $\D^L_N$, finite size effects appear in a completely similar way.
 It also implies from  an experimental point of view that it is preferable
to work with thin wires with a few transverse channels in order to reach
the condition $T_K^0< \D/N$. It is also worth mentioning that while 
in the 
$1$ channel case, speaking about the ratios between $\D/T_K^0$ or $\xi_K^0/d$
were equivalent, it is not the case here. We can no longer simply compare the Kondo screening cloud size to the length of the wire. Nevertheless, we can still
 define an effective length $d_{eff}=Nd$ which can be compared with $\xi_K^0$.\\

\subsubsection{With channel mixing}
Suppose we now work  with a more general model than (\ref{hamgen}) 
with a tunneling  matrix
$t_{WL,\al,\be}^L$ between the left lead and the left wire (and the same in the right lead).
We may express $c_{-1,\al}$ in terms of the  eigenstates, 
$c_{L,\ga}(\epsilon)$ of the left part of $H_L+H_{W}+H_{LW}+H_G$:
\beq
c_{-1,\al} = \int_{-2t-\mu}^{2t-\mu} d\epsilon \sum_\ga f_{L,\al,\ga}(\epsilon)c_{L,\ga}(\epsilon).
\eeq
The local density of states in the channel $\al$ at site $-1$ 
can be then defined by 
$\rho_\al(\eps)=\sum_\ga |f_{L,\al,\ga}(\epsilon)|^2$ and is normalized accordingly. This LDOS has a level spacing of order $\D\sim \hbar v^F_\al/d_L$
provided all the elements of the tunneling matrix are small compared to $t$.
It implies that the resonance peak positions of $|f_{L,\al,\ga}|$
are almost similar $\forall~ \ga$ given by the ones of the isolated wire (up to a small shift
of the order of the peak width which depends on the tunneling matrix).
\begin{widetext}
In this basis, the Kondo Hamiltonian reads:
\bea
H_K&=&\int\int d\eps d\eps' \sum\limits_{\al,be}\sum\limits_{\ga,\de}
\left[
f^*_{L,\al,\ga}f_{L,\al,\ga} J_{\al,\be}^{LL} c^\dag_{L,\ga}(\epsilon)
{\vec \s\over 2}\cdot \vec S c^\dag_{L,\de}(\epsilon')
+f^*_{R,\al,\ga}f_{R,\al,\ga} J_{\al,\be}^{RR} c^\dag_{R,\ga}(\epsilon)
{\vec \s\over 2}\cdot \vec S c_{R,\de}(\epsilon')\right]\nn\\
&+&\left[f^*_{L,\al,\ga}f_{R,\al,\ga} J_{\al,\be}^{LR}
 c^\dag_{L,\ga}(\epsilon){\vec \s\over 2}\cdot \vec S c_{R,\de}(\epsilon')+H.c\right]
,\eea

\end{widetext}
where the couplings $J_{\al,\be}^{i,j}$ have been defined in Eq. (\ref{defJs}).
One may define new tunneling amplitudes $v^{i}_{\ga}=\sum_\al f_{L,\al,\ga}
t_{WD,\al}^i$ (with $i=L,R$) and a new set of Kondo couplings: 
\beq
\tilde \lambda^{ij}_{\ga,\de}(\eps,\eps')=(v^{i}_{\ga})^* (\eps)v^{j}_{\de}(\eps')/\tilde \eps_D.\eeq

In this new basis, the problem takes a more familiar form with a 
$2N\times 2N$ matrix of Kondo couplings (if we again switch to the compact notation 
where no dictinction between left and right is made):
\beq
H_K=\int\int d\eps d\eps'\sum\limits_{\ga,\de=1}^{2N}\tilde \lambda_{\ga,\de} c^\dag_\ga(\eps) {\vec \s\over 2}\cdot \vec S c_\de(\eps')
\eeq
 The RG equations 
have already been derived and read in this notation:
\beq
\tilde \lambda_{\ga \de}\to \tilde \lambda_{\ga \de}+
 \tilde \lambda_{\ga \de} \sum_{\nu}\int d\epsilon  
{\tilde \lambda_{\nu \nu}(\eps)\over |\epsilon |}.\eeq
The coupling $\tilde \lambda_{\nu \nu}(\eps)$ has some non trivial energy
dependence, since its bare value $\tilde \lambda_{\nu \nu}(\eps)\propto \sum_\al |f_{\al,\nu}(\eps)
t_{WD,\al}|^2$. The average resonant peak spacing of $\sum_\al |f_{\al,\nu}|$
 is 
$O(\hbar v_F/Nd)$ and the position of the peaks are almost 
independent of $\nu$. 
Therefore, in order to calculate the Kondo temperature of the system,
we need to compare $T_K^0$ to the energy scale $O(\hbar v_F/Nd)\equiv \D/N$ as
in the non channel mixing case. To summarize, for weak tunneling amplitudes between wires and leads, 
the mixing of the channels just affects the fine structure of the local density
of states seen by the quantum dot (i.e. the width of the peaks) but not the 
resonant peak positions (and therefore the level spacing) which is almost determined by the diagonalization of the isolated finite size wire.
We note that this conclusion is quite general and  should also be valid for the more realistic experimental situation
of finite size wires connected to $3D$ reservoirs.

\section{Study of transport properties}\label{transport}

We  apply in this section the results obtained in the previous section in order to
consider transport properties, in particular, the conductance through the device under consideration. As for thermodynamic properties, a  perturbative calculation is instructive. We expect it to be valid 
at sufficiently high $T\gg T_K$ when the renormalized couplings are sufficiently 
small.
The linear conductance (for $t_{LW}^i\ne 0$) at cubic order in $J,V$  is 
given by:
\begin{widetext}
\bea
G(T) &=&  {e^2\over \pi\hbar}\pi^2\int d\epsilon \rho_L (\epsilon )  \rho_R (\epsilon )
[-dn_F/d\epsilon ]{3\over 4}J_{LR}^2[1+J_{LL}~ I_L(\eps)+J_{RR}~ I_R(\eps)] ,\nn\\
&+&{e^2\over \pi\hbar}4\pi^2V_{LR}^2\int d\epsilon 
\rho_L (\epsilon )\rho_R (\epsilon )[-dn_F/d\epsilon ]\label{Gf}
\eea
\end{widetext}
with 
\beq
I_L(\eps)=\int d\epsilon' PP{\rho_L(\eps')\over
(\epsilon'-\epsilon)}(1-2n_F(\eps'))
\eeq 
and a similar expression for $I_R$. $PP$ stands for principal part and $n_F(\epsilon )$ is the Fermi distribution function at temperature T.  These 
auxiliary functions  leads to the usual logarithmic corrections. The derivation of this formula
is sketched in the appendix \ref{conductance}.
We have not written the other contributions at cubic order 
since they do not involve 
logarithm divergences (and equal zero for particle-hole symmetry) 
and therefore do not 
renormalize in the infrared limit (see appendix \ref{conductance}).

Notice too that the integral in the expression
of  $I_i(\eps)$ ($i=L,R)$ 
depends on
the local density of states $\rho_i(\eps)$.   
 
Let us focus on the second order terms in $J_{LR}$ and $V_{LR}$, and
ignore, for the moment, the corrections of higher order. We 
must distinguish 3 regimes of temperature resulting simply from the 
fact that $(-dn_F/d\epsilon)$ has a peak with width of $O(T)$.  If $T>>\Delta_n^i$, then 
the integral in Eq. (\ref{Gf}) averages over many peaks of 
$\rho_i (\epsilon )$ so that $G$ is approximately independent of $\epsilon_W^i$:
\begin{equation}
G\approx {e^2\over \pi \hbar}(\pi \rho_0)^2[3J_{LR}^2/4+4V_{LR}^2]\label{GH}
\end{equation}
where $\rho_0=\sin k_F/\pi t$ is the average local density of states.

  When $\delta_n^i <<T<<\Delta_n^i$, the conductance 
depends strongly on the gate voltages $\epsilon_W^i$. We then 
encounter three possibilities: i) both $\eps_W^i$ are tuned to a resonance
peak, (ii) both are far from a resonance peak, (iii) only one of them, let us choose
$\eps_W^L$ is tuned to a resonance peak.
Let us detail these three cases:

(i) If both $\epsilon_W^i$ are tuned to a 
resonance peak such that $\epsilon_W^L=\mu+2t\cos[k_{L,n}]$ and
$\epsilon_W^L=\mu+2t\cos[k_{L,m}]$, then 
the integral in Eq. (\ref{Gf}) is dominated by the peaks located 
at $k_{L,n}\app \pi n/(d_L+1)$ and $k_{R,m}\app \pi m/(d_R+1)$
(notice that the integers $n$ and $m$ are in general different).
In this regime of temperature, we can approximate $(-dn_F/d\eps)\approx 1/4T$.
Moreover, we can neglect the other peaks except the resonant ones in the LDOS:
$$\pi \rho_L(\eps)\app {2\sin^2(k_{L,n})\over d_L} {\de_n^L\over \eps^2+(\de_n^L)^2},$$
and we have a similar expression for $\pi \rho_R(\eps)$. Performing the integration we obtain for the on resonance conductance (subscript R):
\begin{widetext}
\bea
G^R(T)&&\approx {e^2\over \pi\hbar}{(3J_{LR}^2/4 + 4V_{LR}^2)\over T d_L d_R}
{\pi \sin^2(k_{L,n})\sin^2 (k_{R,m})\over \de_n^L+\de_m^R}\nn \\
&&\approx {e^2\over \pi\hbar}(3J_{LR}^2/4 + 4V_{LR}^2)
{\pi  t \sin^2(k_{L,n})\sin^2 (k_{R,m})\over 2T\sin k\ [(t_{LW}^L)^2 d_R  \sin^2(k_{L,n})
+(t_{LW}^R)^2 d_L  \sin^2(k_{R,m})]}
\label{gor}
\eea
For a completely symmetric geometry, this expression reads:
\beq
G^R(T)\approx {e^2\over \pi\hbar}{(3J_{LR}^2/4 + 4V_{LR}^2)
\over 2T d t_{LW}^2 \sin k}\pi t^2 \sin^2(k_{i,n})
\eeq
with $k_{i,n}=k_{L,n}=k_{R,m}$.

(ii) On the other hand, if both $\epsilon_W^i$ are far from a resonance peak 
(compared to $T$) then the LDOS is almost  constant and may be 
approximated as
$\pi \rho_L(\eps)\app {2\sin^2(k_{L,n})\over d_L} {\de_n^L\over \D_L^2}$.
A similar
expression for $\rho_R(\eps)$ can be written. 
Therefore the non resonant (subscript NR) conductance reads: 
\bea
G^{NR}(T)&\approx& {e^2\over \pi\hbar}(3J_{LR}^2/4 + 4V_{LR}^2){\sin k_{L,n}\sin k_{R,m}\over\pi^2}  {\de^L_n\de^R_n\over (\D_n^L\D_n^R)^2}\nn\\
G^{NR}(T)&\approx& {e^2\over \pi\hbar}(3J_{LR}^2/4 + 4V_{LR}^2){(t_{LW}^L)^2(t_{LW}^R)^2 
\sin^2 k_{L,n}\sin^2 k_{R,n} \sin^2 k\over \pi^4 t^6}. 
\label{Gmo}
\eea
We also give the expression for the completely symmetric geometry
\beq
G^{NR}(T) \approx {e^2\over \pi\hbar}(3J_{LR}^2/4 + 4V_{LR}^2)
{t_{LW}^4\sin^4(k_{i,n})\sin^2 k\over \pi^4 t^6}.\eeq

We immediately notice that $G^{NR}(T)/G^R(T)=O(t_{LW}^6/T d t^5)\ll 1$.

(iii) Finally, if only $\eps_W^L$ is tuned on resonance, then the conductance associated to this hybrid (H) situation reads:
\bea
G^H(T)&\approx& {e^2\over \pi\hbar}(3J_{LR}^2/4 + 4V_{LR}^2){(t_{LW}^R)^2 
\sin^2 k_{L,n}\sin^2 k_{R,n} \sin k\over 2\pi T t^3 d_L } \nn\\
&\approx&{e^2\over \pi\hbar} (3J_{LR}^2/4 + 4V_{LR}^2){ \D_n^L\over\D_n^R}  {\sin k_{L,n}\sin k_{R,n}\de^R_n\over 4\pi T t^2}.
\label{Gmro}
\eea

The conductance $G^H$ in this hybrid case is also small compared to $G^R$: 
$G^{H}(T)/G^R(T)=O(t_{LW}^4/T d t^3)\ll 1$.

\end{widetext}
In the ultra-low temperature regime, $T<<\delta_n^i$, one may  evaluate  the conductance
in a similar way.
by approximating $\pi\rho_i(\eps)$ by $2\sin^2(k_{i,n})/d_i\de_n^i$. When 
 $\eps_W^L$ and $\eps_W^R$ are both on resonance, we can approximate 
 $\pi\rho_i(\eps)$ by $2\sin^2(k_{i,n})/d_i\de_n^i$. 
The conductance then reads
\begin{equation}
G(T)\approx {e^2\over \pi\hbar}{(3J_{LR}^2/4 + 4V_{LR}^2)t^2\over (t_{LW}^Lt_{LW}^R)^2
\sin^2 k}
\label{Gol}\end{equation}
The conductance is still given by Eq. (\ref{Gmo}) when $\eps_W^L$ and $\eps_W^R$ are both
 tuned off 
 resonance for $T<\de_n$. Finally, when only $\eps_W^L$ is tuned on resonance,
the conductance reads: 
\begin{equation}
G(T)\approx {e^2\over \pi\hbar}((3J_{LR}^2/4 + 4V_{LR}^2){\sin^2  k_{R,n}\over \pi^2}{(t^R_{LW})^2\over t^2(t^L_{LW})^2}
\label{Golr}\end{equation}

We have used perturbation theory in order to evaluate the finite temperature conductance
in all different situations. We have seen that the on-resonance conductance (where both wires are tuned on resonance) is far larger that the conductance in the other situations. This conclusion is valid only where perturbation theory applies.
These approximate 
formulas certainly break down when they do not give $G\pi\hbar /e^2<<1$, due to higher 
order corrections in $J$ and $V$.  
So our approximate formulas will certainly break 
down before $T$ is lowered to $\delta^i_n$ unless $J<<(t^i_{LW})^2/t$, a 
condition which might typically not be satisfied. When these formulas apply, 
we clearly see that the conductance is much larger when $\eps_W^i$ are tuned
on resonance. 
 
However there is another, more interesting reason why these formulas 
can break down at low $T$, namely Kondo physics.  The cubic correction 
in Eq. (\ref{Gf}) contains a $\ln T$ term which essentially replaces 
$J_{LR}$ by its renormalized value at temperature $T$, $J_{LR}^{eff}(T)$
following (\ref{RG}). We expect 
that this will remain true at higher orders.

Now consider the behavior of the conductance as a function of $T$ and
$\epsilon_W^i$ in the two cases. In the case $\xi_K<<d_i$, we may calculate 
the conductance perturbatively in $J_{LR}^{eff}(T)$ at $T>>T_K^0$ and using 
local Fermi liquid theory for $T<<T_K^0$.\cite{Nozieres}  For $T>>T_K^0$, we obtain 
Eq. (\ref{GH}), essentially independent of $\epsilon_W^i$.  
On the 
other hand, for $T<<T_K^0$,  the conductance reduces to that of two  reservoirs
connecting each other by a contact whose conductance is
$$G_W=2e^2/h ~(2t_{WL}^Lt_{WL}^R)^2/(~(t_{WL}^L)^2+(t_{WL}^R)^2)^2$$ 
(the low temperature
 conductance of the quantum dot in the Kondo regime).
 It corresponds to  our original model with $U=0,~
\eps_d=0$,  and some effective length $\tilde d_i\sim d_i$. ($\tilde d_i$
 can be
somewhat reduced from $d_i$ by an amount of order $\xi_K$). A general but less readable
formula for the transmission associated to such non interacting geometry is given in appendix \ref{resonant}. 
When $t_{WL}^L=t_{WL}^R$ and $\eps_W^L=\eps_W^R=\eps_W$, the quantum dot reaches the unitary limit and  
the conductance reduces to that of an ideal wire of length $\tilde L=\tilde d_L+\tilde d_R$. The expression of such conductance reads
\begin{widetext}
\beq
G(T)={e^2\over \pi \hbar}\int 2t\sin k~ dk{4(tt_{LW})^4\sin^2k_d\sin^2 k\over
A^2+B^2}[{-dn_F\over d\eps}(k)]\eeq
with $k_d$ defined by $\cos k_d=\cos k-\eps_W/2t$, and
\bea
A&=&\sin [k_d(2\tilde L+2)]-(\ga_{L}^2+\ga_R^2)\cos k\sin [k_d(2\tilde L+1)]+
\ga_L^2\ga_R^2\cos2k\sin [2k_d\tilde L]\nn\\
B&=&-(\ga_L^2+2\ga_R^2)\sin k\sin [k_d(2\tilde L+1)]
+\ga_L^2\ga_R^2\sin 2k\sin [2k_d]\tilde L\nn
\eea
\end{widetext}
As $T$ is lowered below $\Delta_n^i$ the conductance for a symmetrically connected quantum dot develops peaks 
with spacing of order $\Delta_n^i/2$ ($i=L,R$).  
This is the spacing of peaks in the density 
of states of a wire of length $L=d_L+d_R$, containing no quantum dot.  It 
is half the spacing in the density of states of the model with $J=0$, 
discussed above.  Initially, as $T$ is lowered below $\Delta_n$, 
the peak width is of $O(T)$ and the peak height is  of 
$O(2e^2\Delta_n t_{WL}^2/hTt^2)$.
As $T$ is lowered below $\delta_n$ the peak width becomes of $O(\delta_n )$
and the peak height becomes of $O(2e^2/h)$.
For a non symmetrically connected quantum dot ($t_{WL}^L\ne t_{WL}^R$), 
the halving of the period occurs when the Kondo quantum dot has a large enough conductance. In particular for two symmetric wires, we can prove using Eq. (\ref{transm}) 
that the number of resonant
peaks doubles when the condition
\beq
|R_D|< {2|R_{LW}|\over 1+|R_{LW}|^2}
\eeq
is satisfied. In this equation, we have defined $R_D=(t_{LW}^L)^2-(t_{LW}^R)^2/((t_{LW}^L)^2+(t_{LW}^R)^2)$, the reflection probability of
the dot at low temperature and 
$R_{LW}$ the  reflection probability of the left or right weak link.

 \begin{figure}
\epsfig{figure=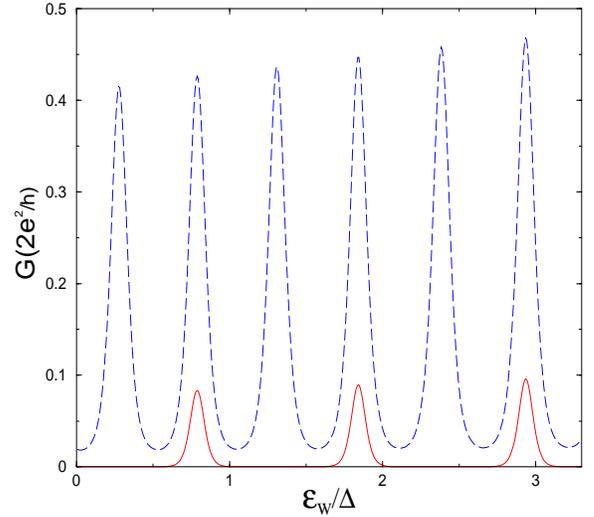,height=7.5cm,width=7cm,angle=-90}
\caption{Conductance in a symmetric geometry as a function of $\eps_W$ at fixed $\D_n$, $T$ and $\de_n$
 for both cases $\xi_K\gg d_L=d_R$
(plain style) and $\xi_K \ll d_L$ (dashed style). We have chosen $T\app \D_n/10\gg
\de_n\app \D_n/100$. The curves have been represented on phase but in general
they are expected to be shifted, the shift being difficult to determine.}
\label{gew}
\end{figure}

On the other hand, when $\xi_K>>d_L,d_R$, the dependence of conductance on $T$ 
and $\epsilon_W$ is very different.  As $T$ is lowered below $\Delta_n^i$ the 
on-resonance conductance starts to grow both because of the single-electron 
effects reflected in  Eqs. (\ref{gor}) and (\ref{Gol}) and, eventually, 
when $T\leq \delta_n$ because of the growth of $J_{eff}(T)$.  However, 
off resonance the conductance stays small, given by Eq. (\ref{Gmo}) 
 at least down to temperatures, 
$T<<\delta_n$ of $O(T_K^{NR})$, given by Eq. (\ref{tkor}). In the temperature 
regime $T_K^{NR}<<T<<\Delta_n$, the conductance has peaks with spacing 
$\Delta_n$ reflecting the fact that $J_{eff}(T)$ is small, off resonance. 
In Figure \ref{gew} we sketch the conductance versus $\epsilon_W$ in the two cases
$T_K^0\gg \D_n$ (dashed style) and $T_K^0\ll \D_n$ (plain style) reflecting the halving of the period between the two curves. Note that we have used perturbation theory
to plot the plain curve explaining why the amplitude is small. Nevertheless, at lower
temperature, 
it is more difficult to calculate the on-resonance 
conductance both because of the breakdown of the perturbative result of 
Eq. (\ref{gor}), (\ref{Gol})  due to single electron effects and because it appears 
considerably more difficult to extract unambiguous predictions from 
local Fermi liquid theory.  Nonetheless, as we will 
check in the next section using  slave boson mean field theory, 
 it is very reasonable to 
expect a conductance of O(1) on resonance at $T\leq \delta_n$   where 
$J_{eff}(T)$ is O(1) on resonance.  Off resonance we can show 
rigorously that the conductance remains small since $J_{eff}(T)$ 
remains small there and so do the single electron corrections 
to Eq. (\ref{Gmo}).  
The 
behavior of the conductance 
 at very low $T\leq T_K^{NR}$ in the case $\xi_K>>d$ 
will be also analyzed using the slave boson mean field theory.

\begin{figure}
\vskip 0.5cm
\includegraphics[height=6cm,width=6cm,angle=-90]{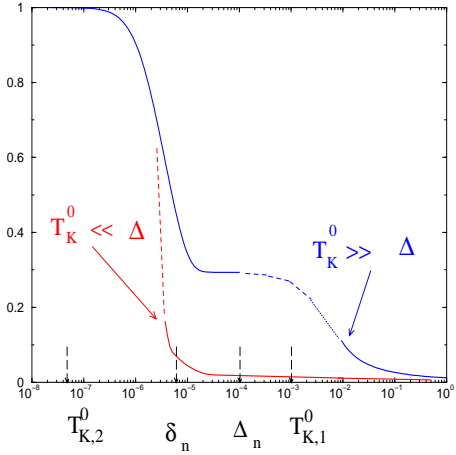}
\caption{Conductance as a function  of temperature in a symmetric device 
(assuming both $\eps_W^i$ is on resonance) for both cases 
$\D_n\ll T_{K,1}^0$ (right blue
curve) and $\D_n\gg \de_n\gg T_{K,2}^0$ (left  red curve). 
The curves in plain style correspond to the perturbative
calculations plus the Fermi liquid result for the first case only. 
We have 
schematically interpolated these curves (dotted lines) where 
neither the perturbative nor the
Fermi liquid theory applies. 
}\label{gtres}
\end{figure}

In Figure \ref{gtres}, we have drawn schematically 
the conductance on resonance as a function of temperature for two different bare Kondo
temperatures $T_{K,1}^0\gg \D_n$ and $T_{K,2}^0\ll \de_n\ll\D_n$, using the
perturbative formula given by Eq. (\ref{Gf}) and the Fermi liquid picture 
valid for the first case only.  For the first case, the conductance has  a plateau which corresponds
to the quantum dot being screened and the $\epsilon$ integral in
Eq. (\ref{Gf}) 
 averaging over many peaks.
The conductance reaches $2e^2/h$ only when $T\ll \de_n$. Conversely, in the 
second case,  the conductance
 remains small til $T\approx \delta$
where the Kondo coupling becomes strongly renormalized (see
Eq. (\ref{jeff})).
We may expect a very abrupt increase of the conductance in this regime as
schematically depicted in Fig \ref{gtres}. Notice that for this choice of
$T_{K,2}^0$, the renormalized Kondo temperature $T_{K,2}^R$ is actually
enhanced  and of order $\de_n$. These different behaviors lead to
different shapes of the curves.

\section{Slave Boson Mean Field approximation}\label{sbmfa}

We will use in this section the Slave Boson Mean Field
Theory (SBMFT) in order to confirm qualitatively our analysis in the previous section.\cite{Hewson}
This approximation seems to well describe qualitatively the behavior of 
the Kondo impurity
at low temperature $T\ll T_K$ when the impurity is screened. 
It has  recently been used by Hu {\it et al.}
\cite{Hu} to calculate the persistent currents in a  metallic ring
containing a quantum dot and their results nicely confirm our perturbative
and renormalization group calculations.
This method assumes $U=\infty$ where the impurity operator 
can be written as $d^\dag_\s=f^\dag_\s b$, where the fermionic operator
$f_\s$ and the bosonic operator $b$ describe the singly occupied electron and holes states respectively.
Furthermore, the constraint $b^\dag b+\sum\limits_\s f^\dag_\s f_\s=1$ has to
be imposed.
In the mean field approximation, the boson operator is replaced by a c-number $b_0$. The constraint is implemented by using the Lagrange parameter $\lam_0$.
Therefore, in the mean field approximation, the Hamiltonian is the same as in (\ref{ham})
with the following changes
\bea
&&H_{WD}=-b_0(t_{WD}^Lc^\dagger_{-1} f+t_{WD}^R f^\dagger_0c_1+H.c.)\\
&&H_{D}=\epsilon_0 f^\dag f+\lam_0(b_0^2-1),\nn
\eea
where we have defined $\eps_0=\eps_D+\lam_0$.

\no Within this approximation, the Hamiltonian reduces to a
non interacting system.
The values of $\lam_0$ and $b_0$ can be determined self-consistently by minimizing the free energy 
of the system defined by $F_{MF}=-{1\over \beta}log Z+\lam_0(b_0^2-1)$
with 
\beq Z=\prod_k(1+e^{-\beta(\eps_k-E_F)})^2\eeq
with $E_F$ the Fermi energy and $\eps_k$ the energy eigenvalues of the mean field Hamiltonian.
Once $b_0$ and $\lambda_0$ are found, we may  directly apply the Landauer
formula to find the conductance at finite temperature of the whole system. 
This is
made possible because the mean field Hamiltonian is a non interacting one.
We also want to emphasize that it is crucial to incorporate from the
beginning the
reservoirs when 
calculating the numerical parameters (here $b_0$ and $\lambda_0$) characterizing
the quantum dot, especially  in the  limit $d_L,d_R\ll\xi_K^0$. 
Note that in [\onlinecite{Thimm}], the quantum dot retarded Green function
was calculated numerically using the non-crossing approximation without
incorporating the reservoirs (i.e. by taking $t_{WL}^i=0$). This might be a crude
approximation especially in the limit $d\ll\xi_K^0$.

The mean field free energy is conveniently expressed as:
\beq
F_{MF}=-{2\over \pi}\int\limits_{-D_0}^{D0} d\eps~ n_F(\eps) \imag (\ln G_f^r(\eps))+\lam_0(b_0^2-1) ,
\eeq
where $n_F(\eps)=1/[1+\exp(\beta(\eps-\mu))]$ and $G_f^r$ is the retarded Green function at the impurity.
Using the approximate expression (\ref{ldosapprox}) for the left and right local density of states, 
the free energy reads:
\begin{widetext}
\bea
F_{MF}={2\over \pi}\int\limits_{-D_0}^{D0} d\eps ~~n_F(\eps) 
\arctan\left( { \sum\limits_{i=L,R}  b_0^2(t_{WD}^i)^2 \ \Sigma_1^i(\eps)\over \eps-\eps_0-
\sum\limits_{i}  b_0^2(t_{WD}^i)^2\ \Sigma_2^i(\eps) }\right)+ \lam_0(b_0^2-1),
\eea
\end{widetext}
where 
\bea
\Sigma_1^i&=& {2\over d_i}\sum\limits_{n=1}^{d_i}\sin^2 k_{i,n}{\delta_{i,n}\over (\eps-\eps_{i,n})^2+
\delta_{i,n}^2 }\\
\Sigma_2^i&=& {2\over d_i}\sum\limits_{n=1}^{d_i}\sin^2 k_{i,n} {(\eps-\eps_{i,n}) \over (\eps-\eps_{i,n})^2+ \delta_{i,n}^2 }.
\eea
By minimizing $F_{MF}$ with respect to $\lam_0$ and $b_0$, we obtain a set of two self consistent equations which can be solved numerically by iteration.
Since the formulation of the problem is non-interacting, the conductance can be expressed using the  Laudauer formula 
\beq
G={2e^2\over h}\int d\eps ({-dn_F\over d\eps}) T(\eps),
\eeq
where $T(\eps)$ is the total transmission probability through the non interactive device in the SBMFT, the quantum dot is modeled by a simple resonant level with $U=0$, $t_{WL}^i\to b_0 t_{WL}^i$ and $\eps_0=\eps_D+\lam_0$.
The expression for $T$ is given in the appendix B by the Eq. (\ref{transmgen}).

\subsection{Analysis of the symmetric geometry}
As in the previous section, we have fixed $d$ (therefore the level spacing $\Delta$) and considered different bare Kondo temperatures corresponding to both limits
$T_K^0\ll \Delta$ and $T_K^0\gg \Delta$, where $\Delta$ is the level spacing 
at the Fermi energy inside the wires.
In the following, we assume that $t_{WD}^L=t_{WD}^R$, $t_{LW}^L=t_{LW}^R=\gamma t$
and $d_L=d_R=d$.

 \begin{figure}
\epsfig{figure=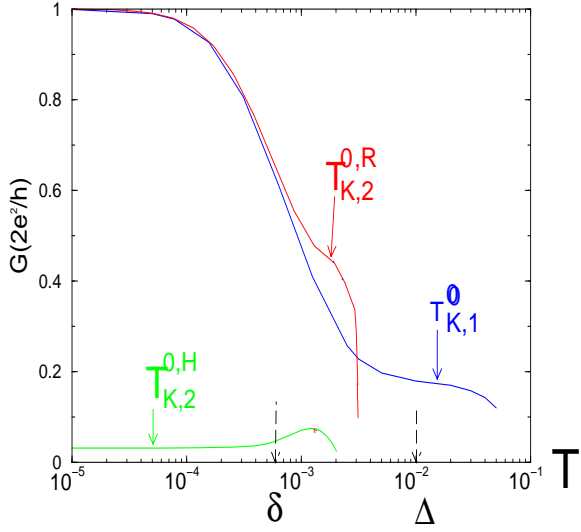,height=7.cm,width=7.5cm,angle=0}
\caption{Conductance from SBMFT as a function of temperature for two bare Kondo temperatures
$T_{K,1}^0\approx 3.6 \D$ and $T_{K,2}^0\approx \D/10$. For the latter case, we have considered the 3 cases depending on $\eps_W^i$ being both on resonance (corresponding to $T_{K,2}^{0,R}$), both off resonance (corresponding to $T_{K,2}^{0,NR}$) or 
the hybrid situation (corresponding to $T_{K,2}^{0,H}$). Here $\gamma=0.3$. We have not plotted the off resonance case because $T_K^{OR}$ seems very small in this case and our SBMFT has convergence difficulties. }
\label{gtgam3}
\end{figure} 

 \begin{figure}
\epsfig{figure=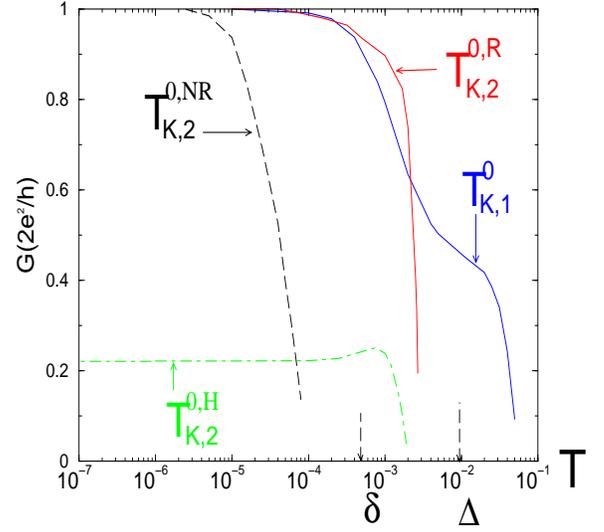,height=7.cm,width=7.5cm,angle=0}
\caption{Same as Fig. \ref{gtgam3} but with $\ga=0.5$. Note that here the 3 cases are plotted for 
$T_{K,2}^0\approx \D/10$.}
\label{gtgam5}
\end{figure} 
We have plotted the conductance as a function of
temperature for two different bare Kondo temperatures $T_{K,1}^0\app 3.6 \Delta$ 
and $T_{K,2}^0\approx \D/10$ for $\ga=0.3$ (Figure \ref{gtgam3}) and $\ga=0.5$ 
(Figure \ref{gtgam5}). For $T_{K,2}^0$, we have considered three cases:
both left and right wires are on resonance, both wires are off resonance and finally only one of them is on resonance. Such a distinction is not worth making for $T_{K,1}^0$ since all curves are very similar since the conductance does not depend of the fine peak structure
of the LDOS. As previously emphasized, the SBMFT is a low temperature method complementary to our perturbation calculation. Therefore, we have stopped the curves at $T\app T_K$
(not the bare one !) where the method failed or is meaningless. Moreover, some convergence
problem may occur in this limit particularly for the case $T_K^0\ll \D$
where the Kondo couplings get very strongly renormalized on a very small energy scale.
Another more powerful numerical method like the Numerical Renormalization Group (NRG)
carried out in Ref. [\onlinecite{Cornaglia}] is required to estimate the conductance for the full crossover from low to high 
temperature.

Let us analyze the results now. In figure \ref{gtgam3}, the curve $T_{K,1}^0$ looks like the one
predicted in Figure \ref{gtres}. It has a plateau around $T=\D$ separating the
low temperature where the unitary limit is reached from the high temperature
perturbative regime (almost not shown here). The curves for $T_{K,2}^{0,R}$ is different
and does not exhibit such an intermediate plateau. On the other hand, when the temperature is increased, it suddenly shoots down abruptly at $T\app T_{K,2}^R\app 0.002\app 3 \de$.
We quickly go from the low temperature regime to the high temperature regime where perturbation theory applies. Note that this sharp drop is also associated with the  limit of convergence
of the SBMFT in this situation (because the variational parameter $b_0$ becomes very small).  This sharp change of behavior in the conductance was predicted by
the perturbative analysis (see Figure \ref{gtres}) and is due to the strong renormalization of the Kondo coupling when the temperature reaches the order of
 the peak width. We have not plotted the conductance for the off resonance case
where our result indicates a Kondo temperature $T_{K,2}^{NR}\app 10^{-7}$ with an abrupt change too. The on resonance- off resonance curve exhibits a similar behavior as the on resonance one but its value at low temperature is very small due to the asymmetry.
For $\ga=0.5$, we still observe an abrupt decrease of the conductance for $T_{k,2}^0$
in the three possible situations. On the other hand, the difference of behavior between $T_{k,1}^0$
and $T_{k,2}^{0,R}$ has almost disappeared . In particular, the plateau around $T\app \D$
for $T_{k,1}^0$ is now replaced by a shoulder. The only way to identify the situation $T_K^0\ll \D$ relies on the dependence on the local densities of states being on or off resonance. 
For this purpose, we  note that the behavior
for the off resonance curve defined by the label $T_{K,2}^{0,NR}$ is very similar to 
the on resonance one defined by $T_{K,2}^{0,R}$, but translated toward lower temperature by roughly two order of magnitudes. 

From this difference, we can immediately infer 
 that the conductance function of the gate voltages $\eps_W^L=\eps_W^R$
exhibits a doubling of the period (compared to the case $T_K^0\gg \D$) in the window
 $T_{K,2}^{NR}\ll T\ll T_{K,1}^0$. Instead of varying two gates voltage, we may imagine
that we fix $\eps_{W}^L$ on resonance and study the conductance as a function of
$\eps_W^R$ only (which is more convenient to realize experimentally). 
Suppose the temperature is also fixed at $T\app \de$ (see Figure \ref{gtgam5}).
The curve corresponding to $T_{K}^0\gg \D$ will exhibit resonance peaks of height $\app 2e^2/h$ with the level spacing $\D/2$ as 
depicted in Figure \ref{gew}. On the other hand, the curve corresponding to 
$T_{K}^0\ll \D$ will exhibit a succession of large resonant peaks(of height $\app 2e^2/h$) and small resonant peaks (of height $\app0.4 e^2/h$), clearly different from the case
$T_{K}^0\gg \D$.

\subsection{Analysis of the complete asymmetric geometry}
 \begin{figure}
\epsfig{figure=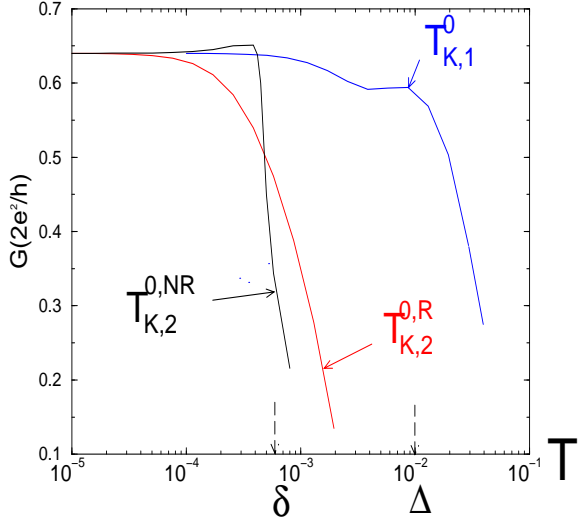,height=7.cm,width=7.5cm,angle=0}
\caption{Conductance as a function of temperature in a completely asymmetric geometry
with only one short wire. We took $\ga_L=0.5$ and $\ga_R=1$. We have considered two bare Kondo temperatures
$T_{K,1}^0\approx 3.6 \D$ and $T_{K,2}^0\approx \D/10$.}
\label{gtass1}
\end{figure} 

 \begin{figure}
\epsfig{figure=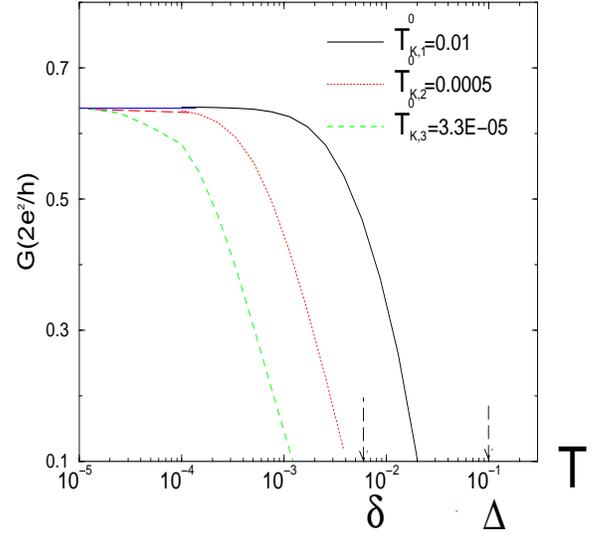,height=7.cm,width=7.5cm,angle=0}
\caption{Same as above. We have considered three  bare Kondo temperatures (see the labels) which all correspond to $T_{K}^0\ll \D$. Note that in this figure $\D\app 0.1$ and $\delta\app 0.006$. } 
\label{gtass2}
\end{figure} 

One may wonder what happens if we connect only one short wire, the left one for example,
 to the quantum wire,
an {\it a priori} easier situation to realize experimentally. In our model, it corresponds to $\ga_R=1$. In the figure \ref{gtass1}, we have therefore plotted the conductance as a function of temperature for 
the two Kondo temperature $T_{K,1}^0=3.6\D\app0.36$ and $T_{K,2}^0=\D/10\app 0.01$
(defined for infinite wires).
Except for  the shoulder around $T\app \D$ for the curve $T_{K,1}^0$, the curves are very much alike. For the geometry with two wires, the best signature of the behavior
$T_K^0\ll \delta$ was the difference between the on- or off- resonance situations. Here this difference is not very dramatic (except for a sharper behavior for $T_{K,2}^{0,NR}$).
The effective Kondo temperatures for both cases are around $T_{K,2}^0=0.001$ (slightly less for the off resonance case). It implies that the renormalization of the Kondo coupling is essentially dominated by the bulk part which does not contain any wire.
In other words, the screening cloud builds mostly in the right lead.
Nevertheless, it may not be always be so. Indeed, we have plotted in 
Figure \ref{gtass2}, the on-resonance conductance for three different  Kondo temperatures $T_{K,1}^0=0.1\D$, $T_{K,2}^0=0.005\D$ and finally $T_{K,3}^0=3.2~10^{-4}\D$.
The first Kondo temperature corresponds to a situation studied in Figure \ref{gtass1} where the screening
is mainly dominated by the right lead. In other words, for such bare Kondo temperature,
we do not expect much variation of the conductance with $\eps_W^L$. The other two
bare Kondo temperatures have been chosen such that $T_{k,2/3}^0 \ll \delta$.
For such choice, we expect that the renormalization of the Kondo couplings will be dominated by the on resonance peak in the left LDOS. This corresponds, as mentioned in the previous section and exemplified here, to an enhancement of the Kondo temperature. 
From the conductance curves, one may infer a cross-over temperature which can interpreted
as an effective Kondo temperature. One could approximately evaluate 
$T_{K,2}\app 0.002 > T_{K,2}^0$ and 
$T_{K,2}\app 0.0005 \gg T_{K,3}^0$. In such cases, one expect a strong dependence of the conductance with $\eps_W^L$. Indeed, when $\eps_W^L$ is off resonance, the renormalization of the Kondo coupling will be again dominated by the right lead with an effective Kondo temperature $T_{K,3}^{OR}\app T_{K,3}^{0}$. It implies that the conductance as a function of $\eps_W^L$ should exhibit peaks of spacing $\D_L$ for the temperature range
$T_{K,3}^{OR}\ll T\ll T_{K,3}^{R}$.
To summarize the analysis of this completely asymmetric geometry, finite size effects
associate with the formation of the Kondo screening cloud may occur only when
the bare Kondo temperature is less that the resonance width. It implied very low Kondo temperatures which may not be within experimental reach.

\section{Discussion and conclusion}\label{discuss}

In this paper, we have presented some detailed (analytical and numerical) 
calculations of a model describing a quantum dot
embedded in two short wires which are connected to reservoirs.
We have shown that finite size effects may occur when the Kondo temperature $T_K^0$ 
(for infinite wires)
becomes of order of the level spacing inside the wires. The main consequence is 
that the genuine Kondo temperature of this system $T_K$ may differ considerably from $T_K^0$ and furthermore 
becomes sensitive to the local fine structure of the local density of states seen by the quantum dot.
The LDOS seen by the dot can be controlled by appropriate gate voltage $\eps_W^{L/R}$.
It is also worth noting that experimentally it provides an extra parameter to control 
the Kondo
temperature. This strong difference between $T_K$ and $T_K^0$ has a signature in all transport and also
thermodynamic properties as we have seen.
An interesting smoking gun to detect these finite size effects is the halving of the period
(see Figure \ref{gew}) in the curve of conductance versus wire gate voltage $\eps_W^{R/L}$.

The symmetric version of the model analyzed in this paper has been investigated recently by Cornaglia and Balseiro \cite{Cornaglia}
using the numerical renormalization group. Their  main conclusions is in agreement
with ours (see also Ref. [\onlinecite{Letter}]).
Nevertheless, on a more quantitative footing, we note that these authors found  a non-monotonic behavior  
in some plots of the  conductance as a function of temperature (particularly for the off resonance case). We have not found
signatures of such behavior. This might be due to the breakdown of the SBMFT
in the regime $T> T_K^{NR}$. We also note that a rather large width of the peaks
in the LDOS was considered by Cornaglia and Balseiro (they typically take 
$\delta\sim T_K^0$) which prevents a distinction between the on-resonance
case for $T_K^0\ll \D$ and the standard case of infinite wires.

The analysis developed in the paper relies on some specific approximations 
we want to
discuss. 

First, we have assumed that all the conductors, the short wires included,
 are in a ballistic regime, or at least that the mean free path $l_d$ is the largest length scale by far. 
If the  mean free path $l_d$ would be for example smaller than
the length of the short wires, we would then need to compare the Kondo length scale
$\xi_K^0$ with $l_d$ and not with $d$ the wire length.

We have also neglected the Coulomb interactions in the wires. 
This is one of the most important approximation made in our treatment
(which also makes the problem tractable).
One open question is whether (and how) the genuine  Kondo temperature of the system 
depends on the Coulomb interactions in the wires.
The Coulomb
interactions may also modify the transport properties through the whole system.
For example, Coulomb blockade effects may appear 
in the conductance as a function of the wire gate voltage. 
Depending on the local density of states being on or off resonance, the conductance
through the central quantum dot will be very large or intead very small (when finite size effects occur).
Assuming Coulomb blockade phenomenon are important in the wires, such a system can be
seen as a double quantum (large) dot connected by a weak/strong tunneling junction,
a device which was studied in Ref. [\onlinecite{Matveev}]. When the tunneling junction is weak
it was found that the period of the peaks doubles compared to the strong tunneling limit.
Indeed, for the small tunneling junction, the excitation spectrum consists of two independent spectra of the two dots/wires of size $L$ whereas for a strong tunneling junction,
the excitation spectrum is better described by the one of a single larger dot of size $2L$.\cite{Matveev} Therefore it  may be tempting to argue that our prediction of the halving of the period in the conductance function of the wires gate voltage still occur when Coulomb interactions are incorporated. 
Nevertheless, a complete  analysis  incorporating consistently 
 both the Coulomb interactions and 
the finite level spacing in the wires would be required and goes belong the scope of 
the present paper. We hope to return to this more difficult but important 
problem in the future.

\vskip 2cm
\appendix
\section{Calculation and analysis of the local density of states}
\label{density}
In this appendix, we give the main steps of the derivation of the expression
for the local density
of states given by Eq. (\ref{rhol})  in our tight binding model. We also  study 
its main characteristics. In the this appendix, we calculate the left LDOS, the right
one is obtained by changing $L\to R$. 

We write the wave function in site $j$ with momentum $k$ as
\bea
\psi(j)&=&A\sin(k_L j) ~~{\rm for}~~ -d_L\leq j\leq -1\nn\\
\psi(j)&=&B\sin(k j)+C \cos(k j) ~~{\rm for}~~ j\leq -d_L-1\nn
\eea
where the wave vectors $k$ and $k_L$ are related through\\
 $-2t\cos k-\mu=-2t\cos k_L +\eps_W^L-\mu$.

\begin{widetext}
$B,C$ are related to $A$ by writing the Schroedinger equations at sites
$-d_L$ and $-d_L-1$:
\bea
B&=& {A\over \ga_L} {\cos (kd_L)\sin(k_L(d_L+1))-\ga_L^2\cos(k(d_L+1))\sin(k_Ld_L)
\over \sin k}\\
C&=& {A\over \ga_L} {-\sin (kd_L)\sin(k_L(d_L+1))+\ga_L^2\sin(k(d_L+1))\sin(k_Ld_L)
\over \sin k }, 
\eea
for $\ga_L=t^L_{WL}/t\neq 0$. Moreover, suppose our system is embedded in a huge box
of size $L>>d$.  Normalizing the wave functions to $1$ in the box
gives the relation   $|B|^2+|C|^2=2/L$.
Therefore, $A^2$ reads
\beq
A^2={2\over L}{\ga_L^2\sin^2 k\over \sin^2(k_L(d_L+1))-2\ga_L^2\cos k\sin(k_L(d_L+1))\sin(k_L \ d_L)+\ga_L^4 \sin^2(k_L\ d_ L)}
\eeq
Notice that for $\ga_L=1$ and $\eps_W^L=0$, we recover $A^2=2/L$.
The local density of states is then given by
$\rho_L(\eps)=A^2sin^2 k_L/(2t\sin k(\pi/L))$
 with our normalization of the anti-commutation relations.
We therefore obtain as a final  expression the equation (\ref{rhol}).

There is another way to derive such expression which is more convenient in order to study the properties of the local density of states.
We consider the same   geometry described in Figure 1. Instead of the quantum dot, we assume 
a direct tunneling  of amplitude $t'$  between the sites $-1$ and $1$.
We also assume the geometry to be completely symmetric (i.e $d_L=d_R$, $\eps_W^L=\eps_W^R$ and $\ga_L=\ga_R$).
For small $t'$ the conductance can be calculated
perturbatively in $t'$:
\beq
G={e^2\over \pi \hbar} 4(t')^2\int d\eps (\pi\rho_L(\eps))^2(-dn_F(\eps)/d\eps)
+O(t'^4)\eeq
Since the system is non interacting,
 the conductance can  be also computed exactly using the Laudauer formula:
\beq
G ={e^2\over \pi \hbar}\int d\eps |T_{tot}|^2(\eps)(-dn_F(\eps)/d\eps)
\eeq
where $T_{tot}(\eps)$ is the total transmission amplitude through the system.
For a wave number $k$, $T(k)$ reads:
\beq
T_{tot}(k)={T_{LW}T_{WL}T_0  \over
  1-2R_0 R_{WL}e^{2ikd_L}+(R_{WL})^2(R_0^2-T_0^2)e^{2ikd}}
\label{transm}
\eeq
where $R_{WL}$ is the reflection coefficient for the left and right weak link
in sites $-d_L$ and $d_R$,
$T_{LW}$ is the transmission coefficient through the weak link for an electron coming from the lead, $T_{WL}$ is the opposite and finally $R_O,T_O$ are the reflection and transmission coefficients through the central weak link.
Since
\beq
T_0(k)={-2it'\sin k_L\over 1-t'^2e^{2ik_L} }~~~;~~~
R_0(k)={t'^2-1\over 1-t'^2 e^{2ik_L} }
\eeq
the term in $t'^2$ is easily obtained by replacing in the denominator
of Eq. (\ref{transm}) $R_0(k)$ by $-1$ and $T_0(k)$ by $0$ and keeping the first term in $T_0(k)$ in the numerator. Therefore, we can write
\beq
\pi \rho_L(k)=\sin k_L
 \left| { T_{LW}T_{WL}  \over
  1-2 r_{wl}e^{2ikd_L+i\pi+i\theta_{WL}}+r_{WL}^2 e^{2ik(d)+2i\theta_{WL}}}\right|
=sin k_L  {t_{lw}t_{wl}  \over
  1-2 r_{wl}\ \cos \al+r_{wl}^2 }\label{rhol2}
\eeq
where we have defined $R_{WL}=r_{wl}e^{i\theta_{WL}}$, $t_{wl}=|T_{WL}|$,
$t_{lw}=|T_{LW}|$ and 
\beq
\al=\pi+2k_L d_L+Arg(R_{WL})=2k_L (d_L+1)+\arctan{\ga_L^2\sin (k+k_L)\over
1-\ga_L^2\cos (k+k_L)}-\arctan{\ga_L^2\sin (k-k_L)\over
1-\ga_L^2\cos (k-k_L)}\label{alpha}.
\eeq
One can check explicitly that the expression in Eq. (\ref{rhol2}) coincides
with the one in Eq. (\ref{rhol}). Fortunately, the equation in (\ref{rhol2})
is much more compact and readable.
One can for example immediately infer from (\ref{alpha}) that the positions
of the peaks in the local density of states are given by the solutions
of $\al=2\pi n$ which gives the equation (\ref{kln}).
The peak half-width $\de_{L,n}$ can also be computed straightforwardly
using the expressions
\beq
r_{wl}=\sqrt{ {(1-\ga_L^2\cos(k-k_L))^2+\ga_L^4\sin^2(k-k_L)\over 
(1-\ga_L^2\cos(k+k_L))^2+\ga_L^4\sin^2(k+k_L) }}~~~;~~~
t_{lw}={2\ga_L\sin k_L\over \sqrt{(1-\ga_L^2\cos(k+k_L))^2+\ga_L^4\sin^2(k+k_L) }}.
\eeq 
The expression for $t_{wl}$ is obtained from $t_{lw}$ by exchanging $k$ with $k_L$.
The height of a given  peak labelled by $n$ is given by 
\beq
\pi  \rho_L(k_n)=\sin k_{L,n} {t_{lw}t_{wl}  \over
  (1-r_{wl})^2 }
\eeq
The width of the peak is obtained for 
$$\de\al\approx 2(d_L+1)\de k_L=(1-r_{wl})/\sqrt{r_{wl}}$$
which provides the expression (\ref{width}) for the width (using $\de \eps=2 \sin k_L \de k_L$).

\section{Conductance through a quantum dot using perturbation theory}
\label{conductance}
In this appendix, we sketch the derivation of the equation (\ref{Gf}).
We follow Ref. [\onlinecite{Glazman}] and first derive a perturbative expression 
for the current assuming there is a small difference of potential $eV=\mu_L-\mu_R$ between the left and right reservoir. The current operator between the left and right side of the dot can be expressed as:
\beq
\hat I(t)={ie\over \hbar}\int d\eps d\eps' ~ f^*_L(\eps)f_R(\eps')\left( J_{LR} c_{L,\eps}^\dagger
{\vec\s\over 2}\cdot \vec S c_{R,\eps'}+V_{LR} c_{L,\eps}^\dagger c_{R,\eps'}-H.c.\right)
\eeq
The current $\langle \hat I \rangle$ can be for example calculated
using
\beq
\langle \hat I \rangle=\langle 0|S(-\infty,0) \hat I(0) S(0,-\infty)|0\rangle
\eeq
with $$S(0,-\infty)=T_K e^{-{i\over \hbar}\int\limits_{-\infty}^0 dt H_{int}(t)}$$
where $H_{int}$ contains the Kondo couplings plus the potential scattering terms
(see Eq. (\ref{HE})). $T_K$ indicates the time ordering operator along a Keldysh contour and $|0\rangle$ designs 
the ground state at time $t=-\infty$.
At second order, we simply pick up the commutator between the current operator and the 
interacting Hamiltonian
\beq
\langle \hat I \rangle^{(2)}=
{2e\over \hbar^2} Re\left\{ \int d\eps d\eps' f^*_L(\eps)f_R(\eps')\int\limits_{-\infty}^0\langle 0|\left[ J_{LR} c_{L,\eps}^\dagger(0)
{\vec\s\over 2}\cdot \vec S c_{R,\eps'}(0)+V_{LR} c_{L,\eps}^\dagger(0) c_{R,\eps'}(0)
,H_{int}(t)\right]|0\rangle\right\}
\eeq
Only the $J_{LR}$ and $V_{LR}$ terms give contributions. After performing the Wick contractions
and taking the small $V$ limit to extract the linear conductance, we recover the second order contributions to 
the expression in Eq. (\ref{HE}).

The calculation of the third order terms goes along the same line. By expanding both $S$ 
operators to second order, we obtain three sorts of terms: a cross term mixing the first order expansions
of $S(-\infty,0)$ and $S(0,-\infty)$ (we denote this term $I_1^{(3)}$) and two other terms coming from the second order expansion of $S(0,-\infty)$ (denoted $I_2^{(3)}$) and from the second order expansion of $S(-\infty,0)$ (denoted $I_3^{(3)}$).

Focusing first on the Kondo terms, the only possible contributions at third order are
in $J_{LR}^2J_{LL}$ and $J_{LR}^2J_{RR}$. Let us calculate for example the terms
in $J_{LR}^2J_{RR}$. 
The mixed term in $J_{RR}$ reads:
\bea
I_{1,R}^{(3)}=&&-{2e\over \hbar^3}J_{LR}^2J_{RR}\int\int\int d\eps d\eps' d\omega~
Im\left\{
\rho^L(\eps)\rho^R(\eps')\rho^R(\omega)\int\int dt_1dt_2 \la S^a(t_1)S^b(0)S^c(t_2)\ra
Tr\left({\s^a\over 2}{\s^b\over 2}{\s^c\over 2}\right) \right. \nn\\
&& \left(  \la c_{L,\eps}(t_1)c_{L,\eps}^\dag(0)\ra
\la c_{R,\eps'}(0)c_{R,\eps'}^\dag(t_2)\ra \la c_{R,\om}^\dag(t_1)c_{R,\om}(t_2)\ra\right.
\\&& \left.\left.
+\la c_{R,\eps'}^\dag(t_1)c_{R,\eps'}(0)\ra \la c_{L,\eps}^\dag(0)c_{L,\eps}(t_2)\ra
 \la c_{R,\om}(t_1)c_{R,\om}^\dag(t_2)\ra \right)
\right\}\nn
\eea
Note that no time ordering is necessary here since the crossed terms belong to 
two different branches of the contour. Using $ \la S^a(t_1)s^b(0)s^c(t_2)\ra={i\over 8}\veps^{abc}$ and $Tr\left({\s^a\over 2}{\s^b\over 2}{\s^c\over 2}\right)={i\over 4}\veps^{abc}$(where $\veps^{abc}$ is the antisymmetric unit tensor), and performing the two time integrations, we can easily evaluate this expression.
The final result reads:
\bea
I_{1,R}^{(3)}=&&{2e\over \hbar}{6\pi\over 32}J_{LR}^2J_{RR}\int d\eps \int d\eps'
\left\{PP {\rho^L(\eps)(\rho^R(\eps'))^2\over \eps-\eps'} n^R_{\eps'}(1-n^R_{\eps'})
\right. \nn\\
&&+\left. PP{\rho^L(\eps)\rho^R(\eps)\rho^R(\eps')\over \eps-\eps'}
( n^R_\eps-n^R_\eps(n^L_\eps+n^R_{\eps'})+n^L_\eps n^R_{\eps'})\right\},
\eea
where $n^{L/R}$ is the occupation number in the left/right reservoir and PP stands for principal part.

We can perform similar calculations for the two other terms denoted $I_{2,R}$ and $I_{3,R}$
paying attention to the different time ordering. We give directly the final expressions for these two terms:
\bea
I_{1,R}^{(3)}=&&-{2e\over \hbar}{6\pi\over 32}J_{LR}^2J_{RR}\int d\eps \int d\eps'
\left\{ PP{\rho^L(\eps)(\rho^R(\eps'))^2\over \eps-\eps'} n^L_\eps n^R_{\eps'}(1-n^R_{\eps'})
\right. \nn\\
&&+\left. PP{\rho^L(\eps)\rho^R(\eps)\rho^R(\eps')\over \eps-\eps'}
n^L_\eps(1-n^R_{\eps})(2-3n^R_{\eps'})\right\},
\eea
and
\bea
I_{1,R}^{(3)}=&&-{2e\over \hbar}{6\pi\over 32}J_{LR}^2J_{RR}\int d\eps \int d\eps'
\left\{ PP{\rho^L(\eps)(\rho^R(\eps'))^2\over \eps-\eps'} (1-n^L_\eps) n^R_{\eps'}(1-n^R_{\eps'})
\right. \nn\\
&&+\left. PP{\rho^L(\eps)\rho^R(\eps)\rho^R(\eps')\over \eps-\eps'}
n^R_\eps(1-n^L_{\eps})(3n^R_{\eps'}-1)\right\},
\eea
Adding this three contributions, we find that the terms in $\rho^L(\eps)(\rho^R(\eps'))^2$
cancel out. The terms in $J_{LR}^2J_{LL}$ can be read out directly by exchanging $L\leftrightarrow R$. For small voltage, we use $n^L(\eps)-n^R(\eps)=eV(-dn/d\eps)$ which provides
the linear conductance given in Eq. (\ref{HE}).

{\it A priori}, one may expect other contributions at third order. The terms in
$V_{LR}^2 J_{LL}$ and  $V_{LR}^2 J_{RR}$ are trivially $0$. 
One can also prove by a similar
calculation that the terms in $J_{LR}^2 V_{LL}$ and $J_{LR}^2 V_{RR}$ 
involve no
logarithm divergencies but instead the integral $PP\int d \eps' \rho^{L/R}(\eps')/(\eps-\eps')$,
an integral which is zero in the continuum limit with $D_0\to \infty$.
The pure scattering contributions in $V_{LR}^2V_{LL}$ and $V_{LR}^2V_{RR}$ 
involve the same integral and therefore do not contain logarithmic divergences
 as it should be. These terms are  negligible in the infrared limit since they do not 
renormalize
as oppposed to the terms involving only the Kondo couplings.

\section{Conductance through a resonant non interacting quantum dot connected to finite size wires}
\label{resonant}

In this appendix, we calculate the conductance through the device depicted in Figure 1 when the quantum dot is modeled by a non interacting resonant level. It corresponds to $U=0$ in (\ref{ham1}).
Generalizing (\ref{transm}) to an asymmetric geometry. the transmission amplitude through the system
\beq
T_{tot}^{lr}(k)={T_{LW}^{lr}T_{WL}^{lr}T_0^{lr}  \over
  1-R_0^l R^r_{LW}e^{2ik_Ld_L}-R_0^r R^l_{WL}e^{2ik_Rd_R}+R_{LW}^rR_{WL}^l
(R_0^lR_0^r-T_0^{lr}T_0^{rl})e^{2ik_Ld_L+k_Rd_R}}
\label{transmgen}
\eeq
where $R_{LW}^r$ is reflection coefficient for the left  weak link (connecting the left lead to the wire) for a wave coming from the right. $R^l_{WL}$ is the opposite. $T_{LW}^{lr}$ is the transmission coefficient through the left weak link for an electron coming from the lead (from left to right), $T_{WL}^{lr}$ is the same for the right weak link, and finally $R_O^r,R_0^l,T_O^{lr},T_0^{rl}$ are the 4 reflection and transmission coefficients through the central weak link.
We write 
\beq
T_0^{lr}=t_0e^{i\vp_0+3i\pi/2}~~~;~~~R_0^r=r_0 e^{i(\vp_0+\tht_0)}~~~;~~~
R_0^l=r_0 e^{i(\vp_0-\tht_0)}
\eeq
with
\bea
t_0^2&=&={4 (t_{WD}^L)^2(t_{WD}^R)^2\sin^2 k_L\over
\left(\eps_0+\cos k_L +\cos k_R -(t_{WD}^L)^2\cos k_L -(t_{WD}^R)^2\cos k_R\right)^2
+\left((t_{WD}^L)^2\sin k_L +(t_{WD}^R)^2\sin k_R\right)^2}\\
r_0^2&=&={ \left(\eps_0+\cos k_L +\cos k_R -(t_{WD}^L)^2\cos k_L -(t_{WD}^R)^2\cos k_R\right)^2
+\left((t_{WD}^L)^2\sin k_L -(t_{WD}^R)^2\sin k_R\right)^2  \over
\left(\eps_0+\cos k_L +\cos k_R -(t_{WD}^L)^2\cos k_L -(t_{WD}^R)^2\cos k_R\right)^2
+\left((t_{WD}^L)^2\sin k_L +(t_{WD}^R)^2\sin k_R\right)^2}
\eea
and
\bea
\vp_0&=&\arctan\left( {(t_{WD}^L)^2\sin k_L +(t_{WD}^R)^2\sin k_R \over
 \eps_0+\cos k_L +\cos k_R -(t_{WD}^L)^2\cos k_L -(t_{WD}^R)^2\cos k_R }\right)\\
\tht_0&=&\arctan\left( {(t_{WD}^L)^2\sin k_L -(t_{WD}^R)^2\sin k_R \over
 \eps_0+\cos k_L +\cos k_R -(t_{WD}^L)^2\cos k_L -(t_{WD}^R)^2\cos k_R }\right)\\
\eea
Using 
\no\beq
T_{LW}^{lr}={2 i \ga_L\sin k\  e^{i(k-k_L)}\over \cos k_L -\ga_L^2 \cos k
-i (\sin k_L +\ga_L^2\sin k) }~~~;~~~
R_{LW}^r=- { \cos k_L -\ga_L^2 \cos k+i (\sin k_L -\ga_L^2\sin k)\over
\cos k_L -\ga_L^2 \cos k-i (\sin k_L +\ga_L^2\sin k) };
\eeq
\no
\beq
T_{WL}^{lr}={2 i \ga_R\sin k_R \ e^{-i(k-k_R)}\over \cos k_R -\ga_R^2 \cos k
-i (\sin k_R +\ga_R^2\sin k) }~~~;~~~
R_{WL}^l=- { \cos k_R -\ga_R^2 \cos k+i (\sin k_R -\ga_R^2\sin k)\over
\cos k_R -\ga_R^2 \cos k-i (\sin k_R +\ga_R^2\sin k) },
\eeq
and the property $R_0^lR_0^r-T_0^{lr}T_0^{rl}=e^{2i\vp_0}$, the expression
(\ref{transmgen}) can be worked out easily. The final result reads
\beq
|T_{tot}|^2={(2t_0\ga_L\ga_R\sin k \sin k_R)^2\over C^2 + D^2}
\eeq
with
\bea
C&=& \cos(k_Ld_L+k_Rd_R+\vp_0)\left( \cos (k_L+k_R)+\ga_L^2\ga_R^2\cos 2k 
-\ga_L^2 \cos k_R \cos k-\ga_R^2 \cos k_L\cos k\right)\nn\\&&
-r_0\cos(k_Rd_R+\tht_0-k_Ld_L)\left( \cos (k_L+k_R)+\ga_L^2\ga_R^2
-\ga_L^2 \cos k_R \cos k-\ga_R^2 \cos k_L\cos k\right)\nn\\&&
-\sin(k_Ld_L+k_Rd_R+\vp_0)\left( \sin (k_L+k_R)-\ga_L^2 \sin k_R \cos k-\ga_R^2 \sin k_L\cos k\right)\\&&
-r_0\sin(k_Rd_R+\tht_0-k_Ld_L)\left( \sin (k_L-k_R)-\ga_L^2 \sin k_R \cos k+\ga_R^2 \sin k_L\cos k\right)\nn
\eea
and
\bea
D&=& -\cos(k_Ld_L+k_Rd_R+\vp_0)\sin k\left(\ga_R^2 \cos k_L+\ga_L^2 \cos k_R-2\ga_R^2\ga_L^2\cos k\right)\nn\\&&
r_0\cos(k_Rd_R+\tht_0-k_Ld_L)\sin k\left(\ga_R^2 \cos k_L-\ga_L^2 \cos k_R\right)\nn\\&&
+\sin (k_Ld_L+k_Rd_R+\vp_0)\sin k\left(\ga_R^2 \cos k_L+\ga_L^2 \cos k_R\right)\\&&
-r_0\sin (k_Rd_R+\tht_0-k_Ld_L)\sin k\left(\ga_R^2 \sin k_L-\ga_L^2 \sin k_R\right)\nn
\eea

\end{widetext}

{\bf Acknowledgments} 
We would like to acknowledge interesting discussions with Carlos Balseiro, 
Claudio Chamon, Leonid Glazman and Mark Kastner.
During the course of this work PS  has been partially supported by the Swiss NSF, 
NCCR, and the EU RTN Spintronics No HPRN-CT-2002-00302. IA was supported by the NSF grant DMR-0203159.


\begin{thebibliography}{999}
\bibitem{dot} D. Goldhaber-Gordon, H. Shtrikman, D. Mahalu, 
D. Abusch-Magder, U. Meirav and M.A. Kaster, Nature {\bf 391}, 156 (1998).
\bibitem{Cronenwett} S.M. Cronewett, T.H. Oosterkamp, L.P. Kouwenhoven, 
Science {\bf 281}, 540 (1998); F. Simmel, R.H. Blick, U.P. Kotthaus,
W. Wegsheider, M. Blichler, Phys. Rev. Lett. {\bf 83}, 804 (1999).
\bibitem{Wiel}
W.G. van der Wiel, S. De Franceschi, T. Fujisawa, 
J.M. Elzerman, S. Tarucha and L.P. Kouwenhoven, Science, {\bf 289}, 2105
(2000).
\bibitem{Cobden} J. Nygard, D. H. Cobden, P. E. Lindelof, 
Nature {\bf 408}, 342 (2000).
\bibitem{Heiblum} Y. Ji, M. Heiblum, D. Sprinzak, D. Mahalu and H. Shtrikman,
Science {\bf 290},779 (2000).
\bibitem{Costi} U. Gerland, J. von Delft, T. A. Costi, and Y. Oreg
Phys. Rev. Lett. 84, 3710 (2000). 
\bibitem{unitarity} O. Entin-Wohlman, A. Aharony, Y. Imry, Y. Levinson, and A. Schiller
Phys. Rev. Lett. 88, 166801 (2002); A. Aharony, O. Entin-Wohlman, B. I. Halperin, and Y. Imry, Phys. Rev. B 66, 115311 (2002)   
\bibitem{Affleck} I. Affleck and P. Simon, Phys. Rev. Lett. {\bf 86}, 2854
  (2001).
\bibitem{Simon} P. Simon and I. Affleck, Phys. Rev. {\bf B64}, 085308 (2001).
\bibitem{Hu} H. Hu, G.-M. Zhang and Yu Lu, Phys. Rev. Lett. {\bf 86}, (2001).
\bibitem{Johan} H.-P. Eckle, H. Johannesson, and C. A. Stafford
Phys. Rev. Lett. {\bf 87}, 016602 (2001)  
Phys. Rev. Lett. {\bf 82}, 5088 (1999).
\bibitem{Kang1} K. Kang and S.-C. Shin, Phys. Rev. Lett. {\bf 85}, 5619 (2000).
\bibitem{Ferrari} V. Ferrari, G. Chiappe, E.V. Anda and M.A. Davidovich, 
Phys. Rev. Lett. {\bf 82}, 5088 (1999).
\bibitem{Aligia} A. A. Aligia, Phys. Rev. {\bf B 66}, 165303 (2002) 
\bibitem{Comment} I. Affleck and P. Simon, Phys. Rev. Lett. {\bf 88}, 139701 (2002); 
H.-P. Eckle, H. Johannesson, and C. A. Stafford, Phys. Rev. Lett. {\bf 88}, 139702 (2002).
\bibitem{Sorensen} E. Sorensen and I. Affleck, unpublished. 
\bibitem{Thimm} W. B. Thimm, J. Kroha and J. von Delft, Phys. Rev. Lett. {\bf
    82} 2143 (1999).
\bibitem{Balseiro} P. S. Cornaglia and C. A. Balseiro, Phys. Rev. {\bf B66}, 115303 (2002); ibid {\bf B66}, 
174404 (2002).
\bibitem{Kang} K. Kang and L. Craco, Phys. Rev. B {\bf 65}, 033302 (2002). 
\bibitem{Letter} P. Simon and I. Affleck, Phys. Rev. Lett. {\bf 89} 206602 (2002).
\bibitem{Cornaglia} P. S. Cornaglia and C. A. Balseiro, cond-mat/0212119.
\bibitem{Cho} S. Y. Cho, H-Q. Zhou and R. H. McKenzie, cond-mat/0302090.
\bibitem{Ng}T. K. Ng and P. A. Lee, Phys. Rev. Lett. {\bf 61} 1768 (1988).
\bibitem{Langreth}D. C. Langreth, Phys. Rev. {\bf 150} 516 (1966).
\bibitem{Glazman} A. Kaminski, Yu. V. Nazarov and L. I. Glazman,
  Phys. Rev. {\bf B 62}, 8154 (2000).
\bibitem{Hewson}, A. C. Hewson, {\it The kondo Problem to Heavy Fermions} (Cambridge University Press, Cambridge, UK, 1997), chap 7.
\bibitem{Nozieres} P. Nozi\`eres,  
{\it Proceedings of the 14th International Conference  
on Low Temperature Physics}, (ed. M. Krusius and M. Vuorio, North-Holland,  
Amsterdam, 1975), Vol. 5, p. 339.
\bibitem{Raikh} S. Kettemann and M. E. Raikh, Phys. Rev. Lett. {\bf 90} 146601 (2003). 
\bibitem{Matveev} K. A. Matveev, L . I. Glazman and H. U. Baranger, Phys. Rev. 
{\bf B 53} 1034 (1996); ibid {\bf B 54} 5637 (1996).
\end{thebibliography}
\end{document}